\documentclass[12pt]{article}
\usepackage{amsmath}
\usepackage{amsfonts}
\usepackage{graphicx,psfrag,epsf}
\usepackage{enumerate}
\usepackage{caption}
\usepackage{subcaption}
\usepackage{natbib}
\usepackage{hyperref}
\usepackage{dsfont}
\usepackage{float}
\usepackage{bbold}
\usepackage{multirow}
\newcommand{\blind}{0}
\usepackage{xcolor}
\usepackage[normalem]{ulem}
\usepackage{mathtools}
\usepackage{comment}
\usepackage{booktabs}

\begin{document}

\def\spacingset#1{\renewcommand{\baselinestretch}%
{#1}\small\normalsize} \spacingset{1}

\if0\blind
{
  \title{\bf A survival analysis of glioma patients using topological features and locations of tumors}
  \author{Yuhyeong Jang$^{1}$, Tu Dan$^{2}$, Eric Vu$^{1}$, and Chul Moon$^{1}$\thanks{
    chulm@smu.edu}\\
    $^{1}$Department of Statistics and Data Science\\Southern Methodist University \\
    $^{2}$Department of Radiation Oncology\\The University of Texas Southwestern Medical Center
    }
  \date{}
  \maketitle
} \fi

\if1\blind
{
  \begin{center}
    {\LARGE\bf Title}
\end{center}
  \medskip
} \fi

\bigskip
\begin{abstract}
Tumor shape plays a critical role in influencing both growth and metastasis. We introduce a novel topological radiomic feature derived from persistent homology to characterize tumor shape, focusing on its association with time-to-event outcomes in gliomas. These features effectively capture diverse tumor shape patterns that are not represented by conventional radiomic measures. To incorporate these features into survival analysis, we employ a functional Cox regression model in which the topological features are represented in a functional space. We further include interaction terms between shape features and tumor location to capture lobe-specific effects. This approach enables interpretable assessment of how tumor morphology relates to survival risk. We evaluate the proposed method in two case studies using radiomic images of high-grade and low-grade gliomas. The findings suggest that the topological features serve as strong predictors of survival prognosis, remaining significant after adjusting for clinical variables, and provide additional clinically meaningful insights into tumor behavior.

\end{abstract}

\noindent%

\spacingset{1.45}

\section{Introduction}
\label{sec:intro}

The rapid integration of Artificial Intelligence (AI) into medical imaging has transformed the way complex diseases such as cancer are analyzed and interpreted. Deep learning models, in particular, have demonstrated impressive capabilities in segmenting lesions and tumors from radiologic and histopathologic images with expert-level performance~\citep{litjens2017survey, esteva2021deep}. These advancements have expanded opportunities for computational biomarker discovery and image-based prognostic modeling.

One growing field at the intersection of imaging and analytics is radiomics that extracts quantitative features, such as intensity, texture, and shape, from medical images to support diagnosis and outcome prediction~\citep{aerts2014decoding, lambin2017radiomics}. Texture features, like entropy and co-occurrence matrices~\citep{haralick2007textural}, offer information on local intensity variation, while shape descriptors often focus on tumor boundary irregularities~\citep{zwanenburg2020reliability}. Despite their success, conventional radiomic features tend to oversimplify tumor geometry or ignore higher-order spatial relationships, limiting their interpretability and robustness \cite{limkin2019complexity,lambin2017radiomics, demirciouglu2025reproducibility}.

Topological Data Analysis (TDA) has emerged as a powerful approach to overcome these limitations by studying the ``shape'' of data in a mathematically rigorous and transformation-invariant manner~\citep{carlsson2009topology}. A central tool in TDA is persistent homology, which captures multiscale topological features, such as connected components, loops, and voids, in data-derived structures~\citep{ghrist2008barcodes, chazal2017introduction}. Persistent homology is inherently robust to noise and does not rely on predefined geometric assumptions, making it well-suited for analyzing complex tumor morphologies. Recent work has applied persistent homology to tumor segmentation~\citep{qaiser2016persistent}, Gleason score clustering~\citep{lawson2019persistent}, histological subtype classification~\citep{oyama2019hepatic}, and bone microstructure quantification~\citep{pritchard2023persistent}. Also, several studies have
focused on survival prediction using topological features of medical images. \cite{Crawford2020} introduce smooth Euler characteristics to characterize tumor boundary shapes and develop a corresponding functional survival model. \cite{somasundaram2021persistent} demonstrate that persistent homology features extracted from Computed Tomography (CT) images can predict survival outcomes in lung cancer patients. More recently, \cite{Moon2023} propose an interpretable survival model tailored to imaging data from lung and brain cancer patients, but they only analyze 2D image data.

The World Health Organization (WHO) classification system categorizes primary brain tumors, from grade \MakeUppercase{\romannumeral 1} to grade \MakeUppercase{\romannumeral 4} based on microscopic histologic characteristics such as cytologic atypia, anaplasia, mitotic activity, microvascular proliferation, and necrosis \citep{Melhem2022Updates, Louis2021WHO}. WHO grade \MakeUppercase{\romannumeral 4} gliomablastoma is the most aggressive form of glioma with poor survival outcomes, exhibiting high-grade histologic features including necrosis and microvascular proliferation \citep{Melhem2022Updates}. In contrast, WHO grade \MakeUppercase{\romannumeral 1}-\MakeUppercase{\romannumeral 2} low grade glioma encompasses a collection of diverse lower-grade tumors that lack the aforementioned aggressive histologic features and is often driven by single driver mutations in the Mitogen-Activated Protein (MAP) kinase or Isocitrate Dehydrogenase (IDH) pathways \citep{Forst2014LowGrade}. LGGs typically occur in younger individuals and follow a more indolent course with substantially longer survival times compared with high-grade gliomas \citep{RAVANPAY2018573}.

In this study, we develop a new framework for modeling lobe-specific effects of topological features extracted from 3D AI-segmented glioma images using persistent homology. By applying persistent homology to distance transforms of segmented masks, we capture a rich set of shape characteristics that are tailored to radiomic images. To connect these features to clinical outcomes, we combine a functional representation of persistence diagrams and the functional Cox regression model, which leads to the persistent homology functional Cox (PH-FCox) model. Our approach leverages Functional Principal Component Analysis (FPCA) to represent results of persistent homology in a lower-dimensional, interpretable form that facilitates survival analysis. Furthermore, the model incorporates interaction terms between topological features and the cerebral locations of tumors. This allows us to flexibly model lobe-specific effects of topological features together with clinical covariates within a principled statistical framework. We conduct a simulation study to demonstrate that topological shape features effectively capture systematic shape variations and can be integrated into statistical modeling via the PH-FCox framework, particularly for risk stratification. 

We apply the proposed model to two types of brain tumors, glioblastoma (GBM) and low-grade glioma (LGG), using 133 and 107 segmented 3D brain MRI scans, respectively. The comparative performance of the PH-FCox model is examined in relation to three benchmark models to evaluate its predictive advantage. 
In addition, we provide interpretations of the model results and their potential clinical implications. This predictive analysis is expected to offer new insights into the prognostic relevance of tumor morphology, with particular attention given to deriving clinically meaningful interpretations.

The remainder of the paper is organized as follows: Section~\ref{sec:methods} describes the construction of persistent homology features and the PH-FCox model. Section~\ref{sec:simul} presents simulation studies on robustness to shape perturbations. Section~\ref{sec:analysis} demonstrates the method on clinical imaging datasets. Section~\ref{sec:conc} concludes with a discussion of implications and future work.

\section{Methodology}
\label{sec:methods}

\subsection{Topological shape feature extraction using persistent homology}
\label{subsec:PH}

We implement the topological data analysis approach based on homology to characterize the shapes of brain tumor images. Homology describes an object in terms of its topological features or holes: 0-dimensional holes (connected components), one-dimensional holes (loops enclosing an empty region), and 2-dimensional holes (voids bounded by a closed surface) \citep{Wasserman2018}. Intuitively, one-dimensional holes can be seen in objects like a rubber band or a ring, while 2-dimensional holes appear as enclosed cavities, such as a soccer ball. These features or holes are invariant under homeomorphism, capturing the essential topological structure of an object rather than its specific geometry. 

Persistent homology extends classical homology by examining how topological features evolve across multiple scales of the data \citep{Hensel2021}. As the scale, called the filtration, increases, new holes may appear, and existing holes may merge or disappear. Persistent homology tracks births and deaths of these features across the filtration, resulting in a multi-scale summary of the tumor’s shape. The output of persistent homology for the $j$-th dimension can be compactly summarized by a persistence diagram, a collection of points in the plane $\mathbb{R}^2$ defined as ${P^{(j)}} = \{(b, d) \in \mathbb{R}^2 \mid b < d\}$. Each point $(b,d)$ indicates a topological feature that appears at filtration value $b$ (birth) and disappears at filtration value $d$ (death).

\begin{figure}[!ht]
    \centering
    \begin{subfigure}[t]{0.45\textwidth}
        \centering
        \includegraphics[height=3cm]{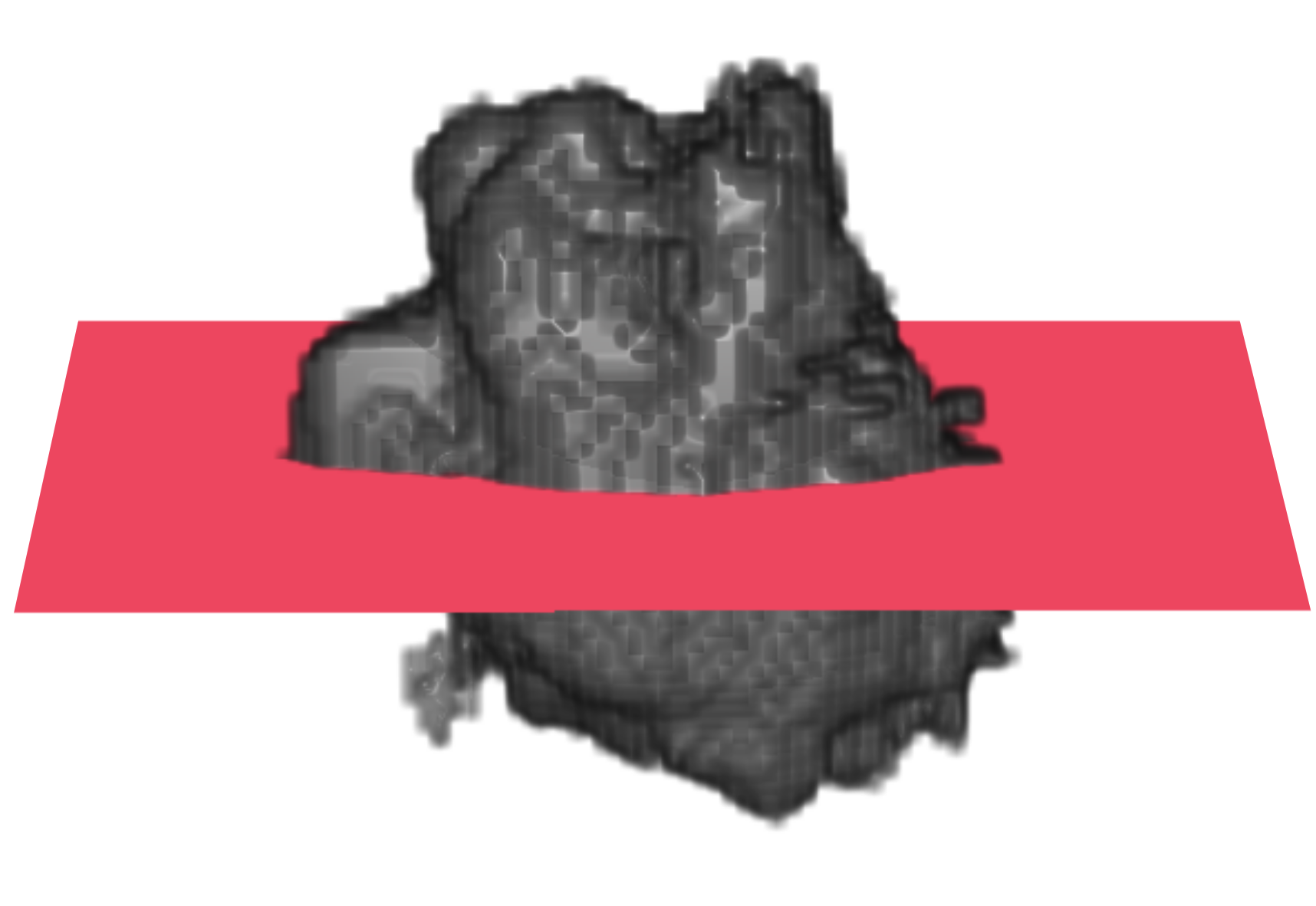}
        \caption{} 
        \label{fig:tumor3d}
    \end{subfigure}
    \begin{subfigure}[t]{0.45\textwidth}
        \centering
        \includegraphics[height=3cm]{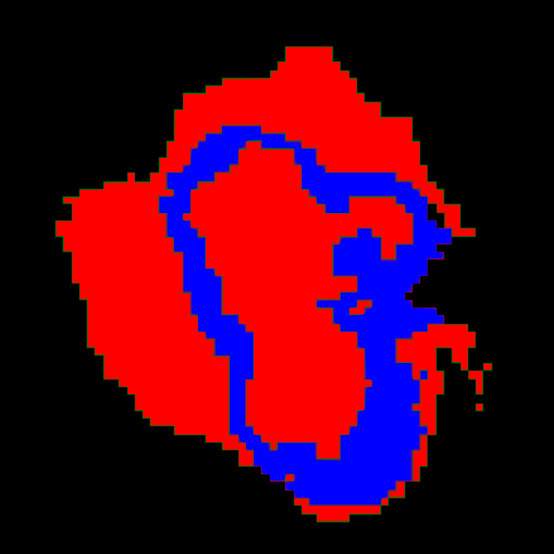}
        \caption{} 
        \label{fig:tumor_three_class}
    \end{subfigure}
    
    \begin{subfigure}[t]{0.45\textwidth}
        \centering
        \includegraphics[height=3cm]{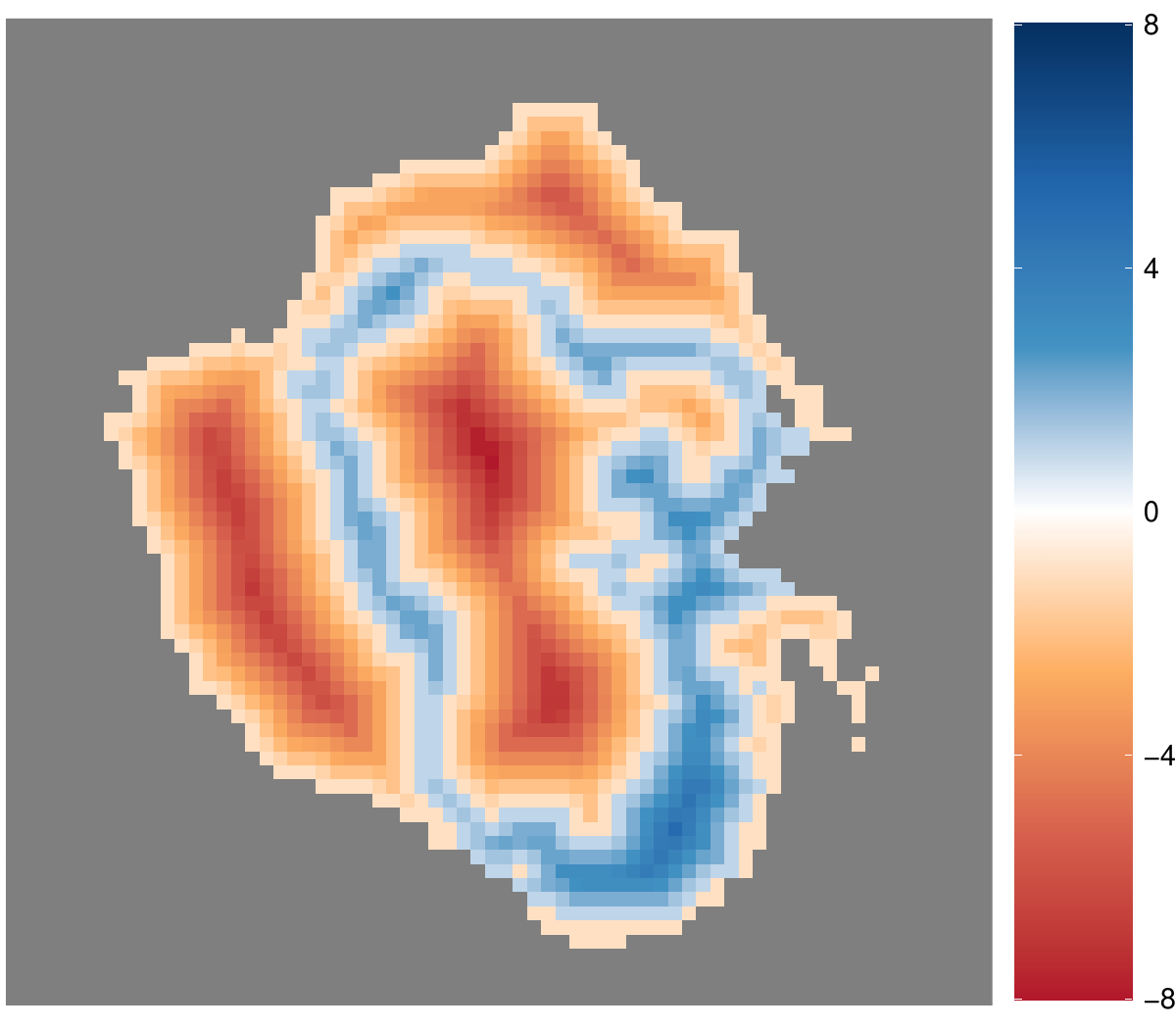}
        \caption{} 
        \label{fig:tumor_sedt}
    \end{subfigure}
    \begin{subfigure}[t]{0.45\textwidth}
        \centering
        \includegraphics[height=3cm]{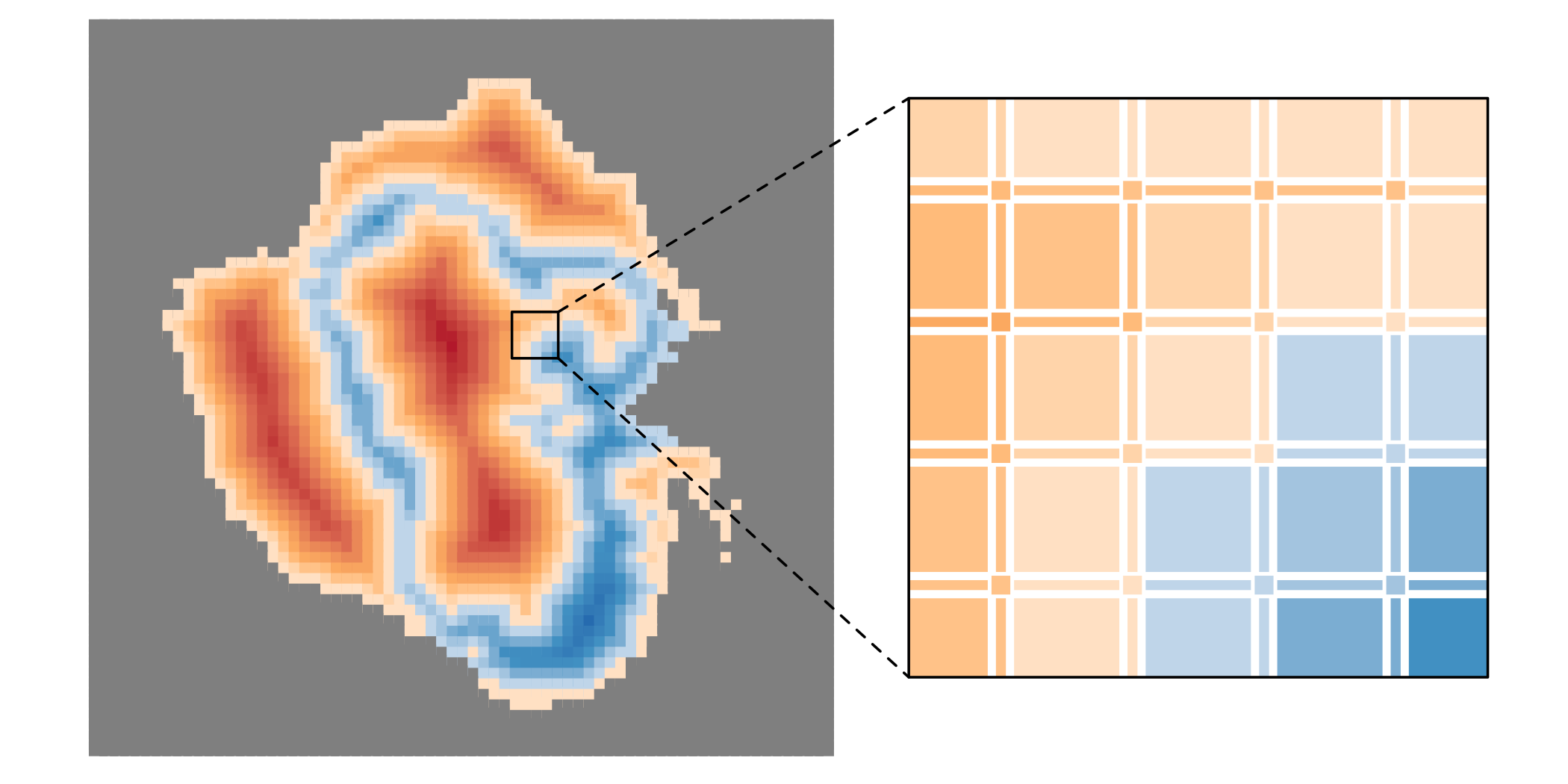}
        \caption{} 
        \label{fig:tumor_cubical}
    \end{subfigure}
\caption{Example of brain tumor image: (a) 3D brain tumor image and the example slice; (b) A slice of a three-class segmented tumor image, cut along the red plane in the 3D plot. The AT, non-AT, and non-tumor regions are colored as blue, red, and black, respectively; (c) computed SEDT-3 values with the black pixels having infinity.}
\label{fig:example_tumor}
\end{figure}

In our study, we aim to analyze three-class brain tumor images whose voxels belong to one of the three distinct labels as determined by Brain Tumor Segmentation (BraTS) dataset \citep{bakas2018identifying}: 1) active tumor (AT) region 
2) non-active tumor (non-AT) region, including necrotic and non-active tumor core (NCR/NET) and peritumoral edema (ED), and 3) non-tumor region. The example 3D and 2D labeled brain tumor images are provided in Figures~\ref{fig:tumor3d} and \ref{fig:tumor_three_class}.

The voxels in these images are merely discrete labels and lack intrinsic shape information, such as size or relative arrangement with other labels. To retrieve shape information from the images, we apply the signed Euclidean distance transform for three-class images (SEDT-3) \citep{Moon2023}. The SEDT-3 assigns to each voxel its Euclidean distance to the nearest voxel of a different label, and a sign is determined according to its label: AT $(-)$, non-AT $(+)$, and non-tumor $(\infty)$ voxels. Figure~\ref{fig:tumor_sedt} presents the SEDT-3 values of Figure~\ref{fig:tumor_three_class}. The signs retain the existing label information, whereas the distance values provide a quantitative measure of how deeply each voxel lies within its region.

The SEDT-3 values are defined on a regular grid, making the voxel domain naturally suited for representation as a cubical complex. This structure provides an organized framework for computing the homology and thereby characterizing the shape features encoded by the SEDT-3.
A filtered cubical complex $\mathcal{C}$ is constructed combinatorially by gluing elementary cubes: vertices (0-cubes), edges (1-cubes), squares (2-cubes), cubes (3-cubes), and their higher-dimensional counterparts \citep{Kaczynski2004}. Following the rule of construction outlined in Section~S.1 of the Supplementary Material, each element is assigned its intensity value.
Figure~\ref{fig:tumor_cubical} presents the cubical complex of the example SEDT image. More details on cubical complexes can be found in \cite{Kaczynski2004}.

\begin{figure}[!ht]
  
    \centering
    \begin{subfigure}[t]{0.19\textwidth}
        \centering
        \includegraphics[height=2.5cm]{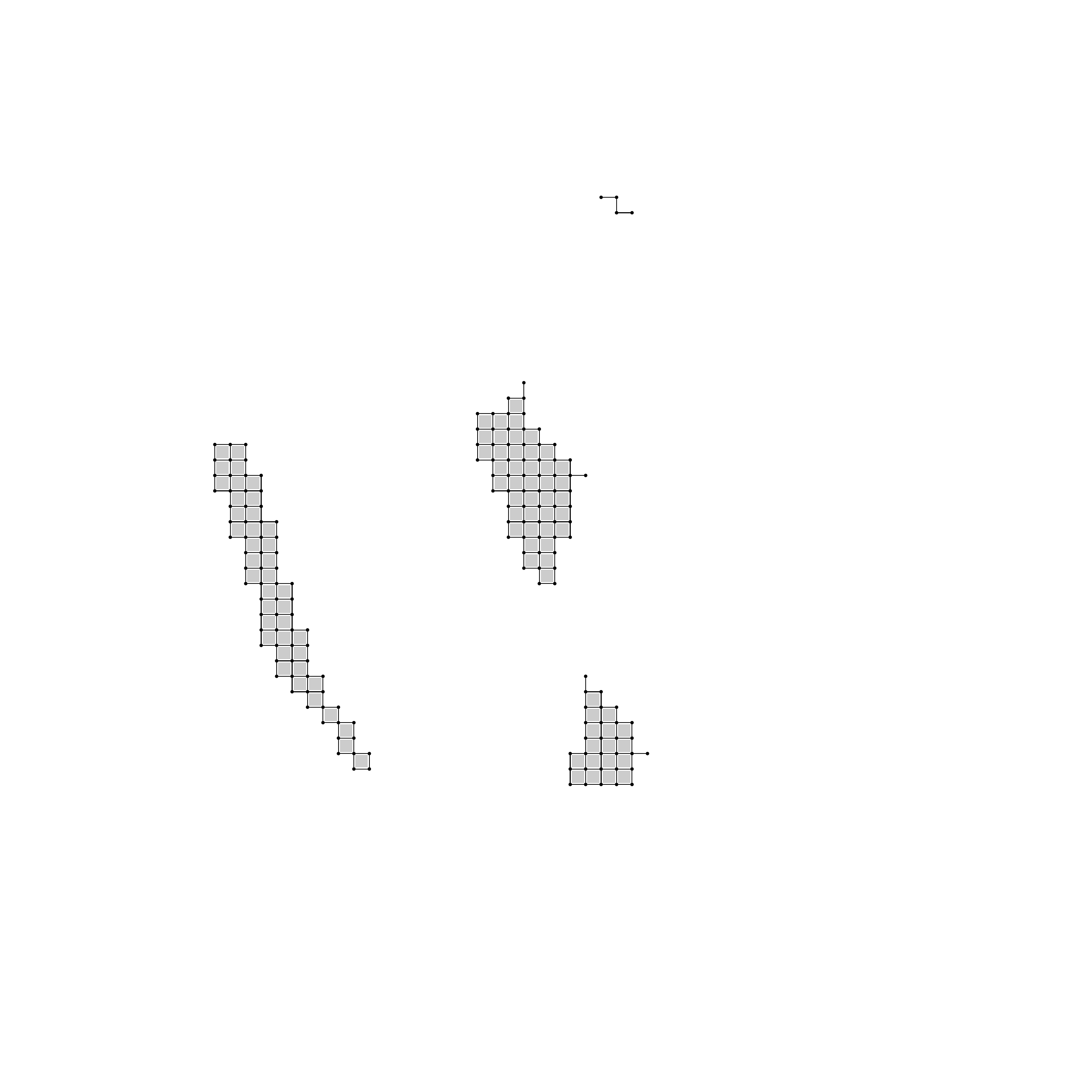}
        \caption{$\epsilon=-5$}
        \label{fig:filt-5}
    \end{subfigure}
    \begin{subfigure}[t]{0.19\textwidth}
        \centering
        \includegraphics[height=2.5cm]{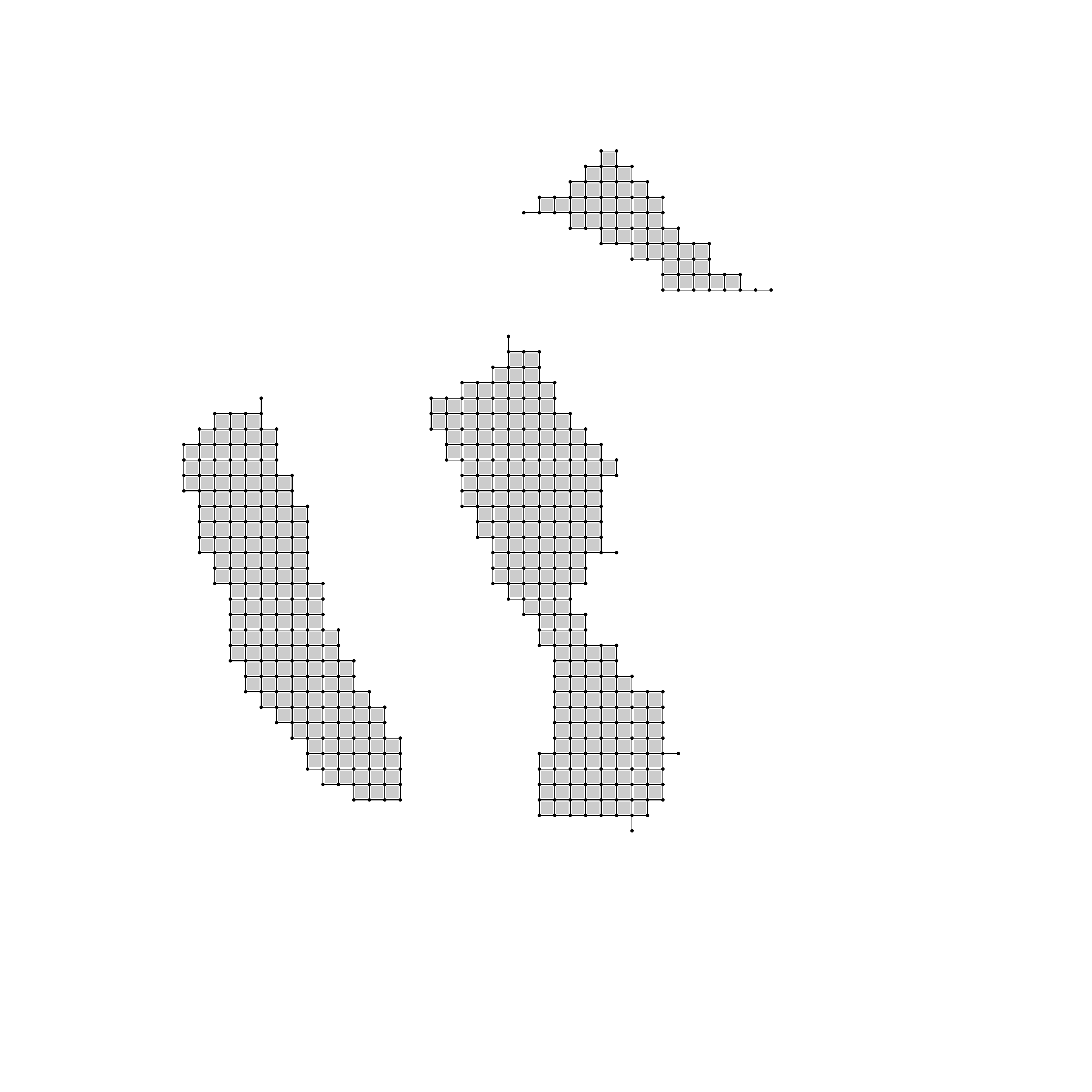}
        \caption{$\epsilon=-3$}
        \label{fig:filt-3}
    \end{subfigure}
    \begin{subfigure}[t]{0.19\textwidth}
        \centering
        \includegraphics[height=2.5cm]{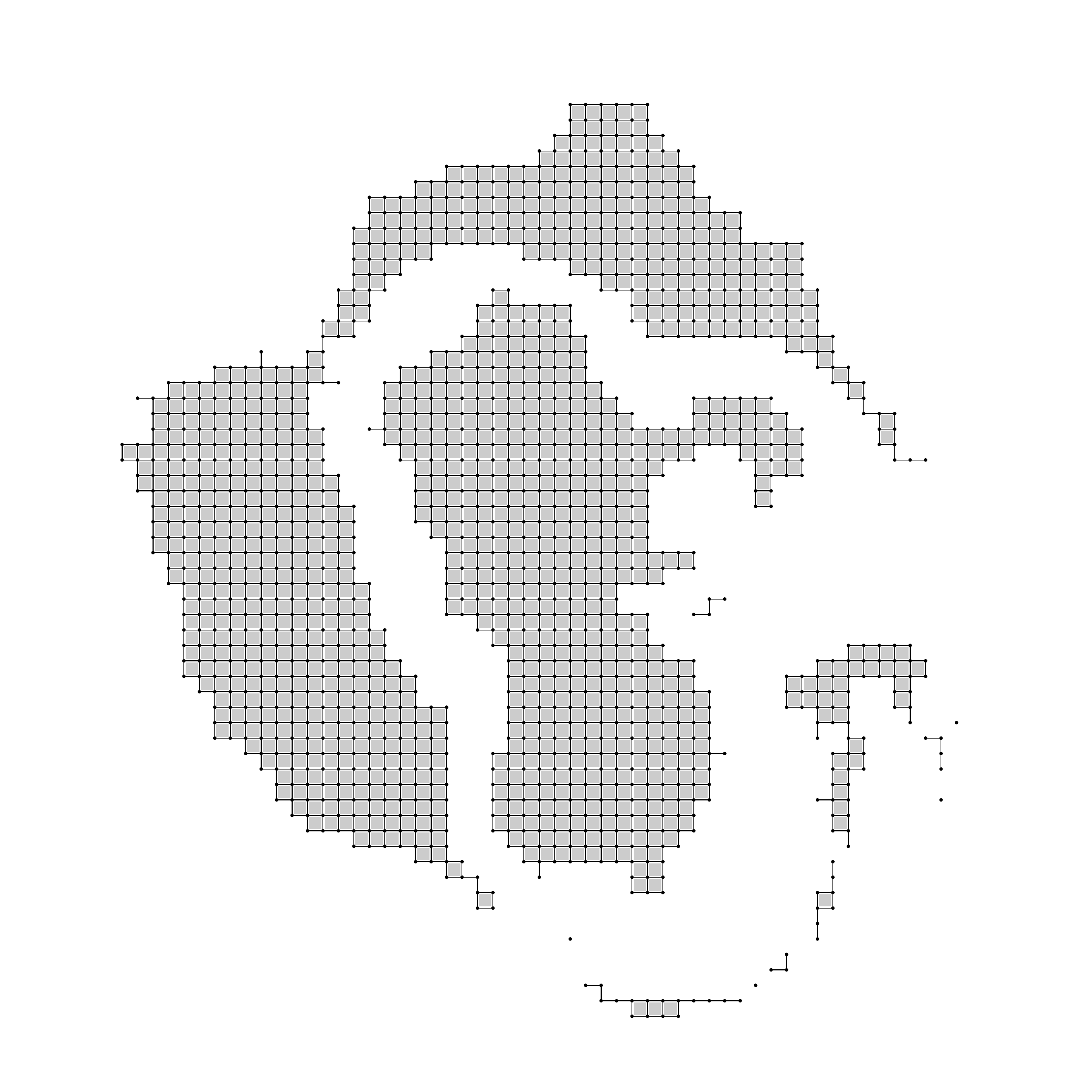}
        \caption{$\epsilon=-1$}
        \label{fig:filt-1}
    \end{subfigure}
    \begin{subfigure}[t]{0.19\textwidth}
        \centering
        \includegraphics[height=2.5cm]{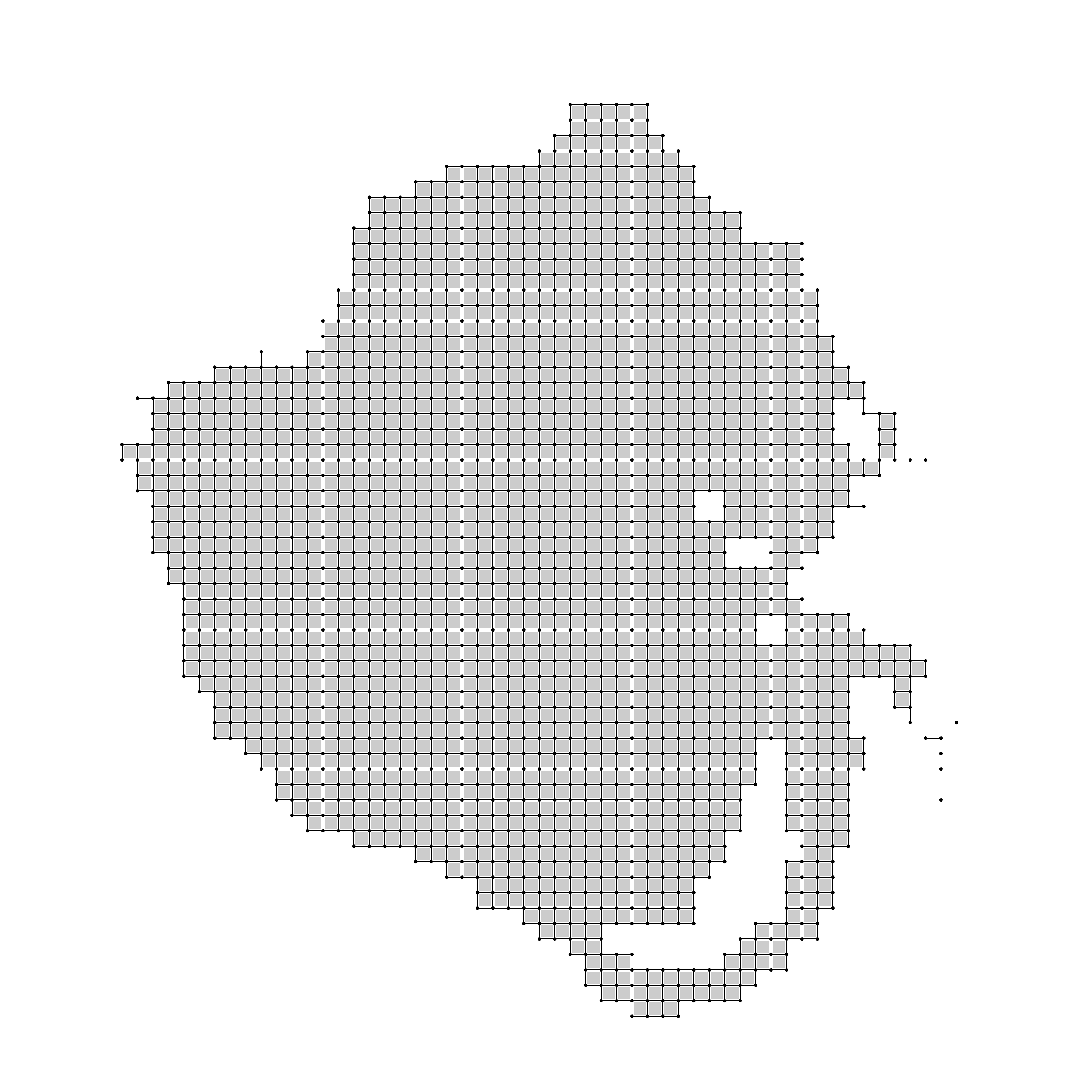}
        \caption{$\epsilon=3$}
        \label{fig:filt3}
    \end{subfigure}
    \begin{subfigure}[t]{0.19\textwidth}
        \centering
        \includegraphics[height=2.5cm]{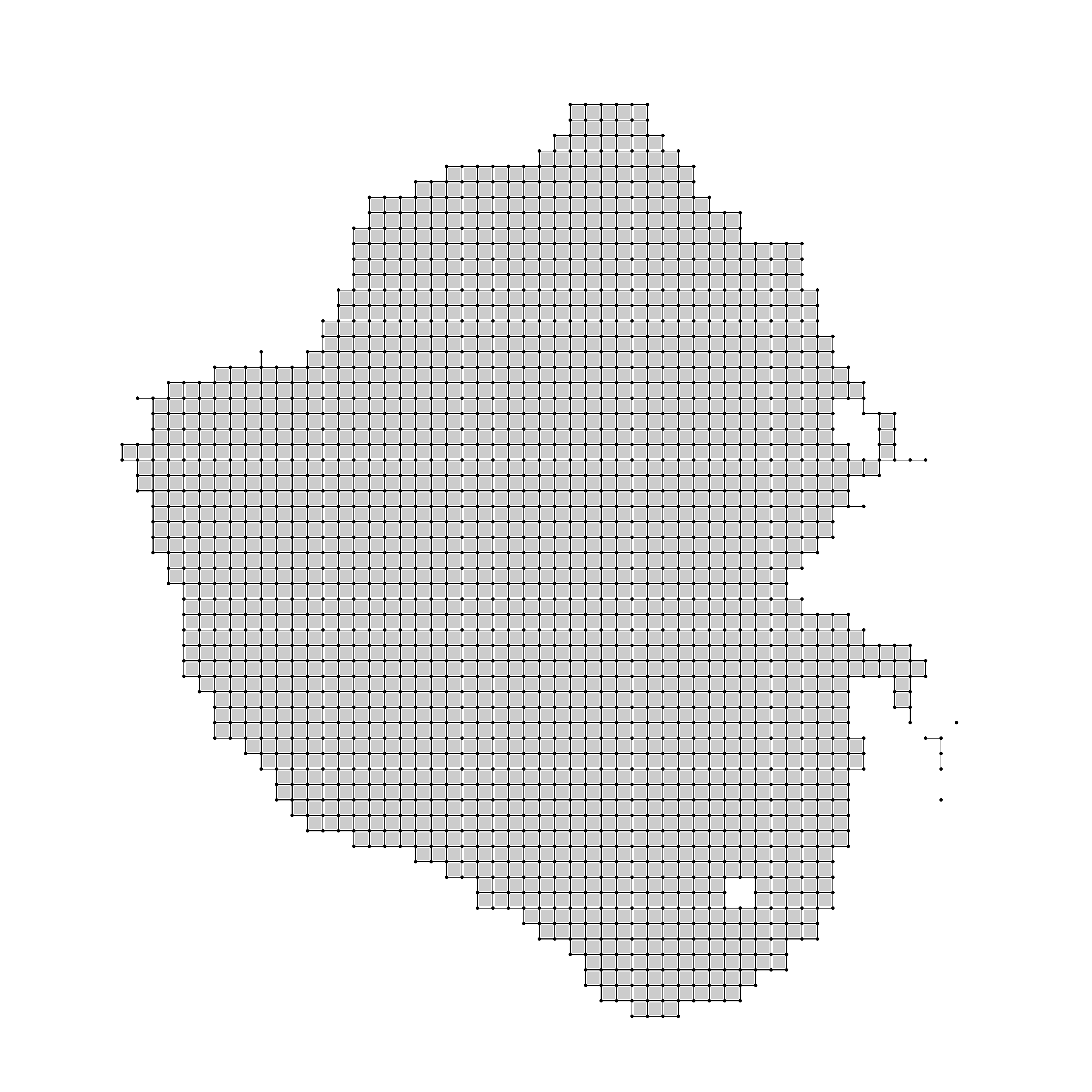}
        \caption{$\epsilon=5$}
        \label{fig:filt5}
    \end{subfigure}

    \begin{subfigure}[t]{0.32\textwidth}
        \centering
        \includegraphics[width=4cm]{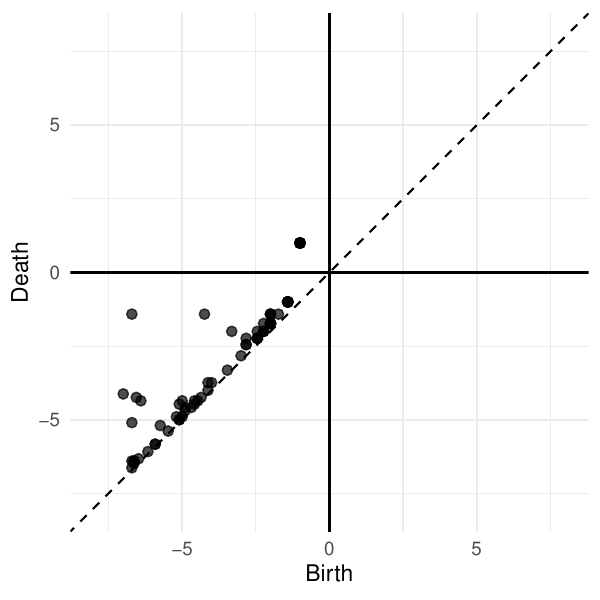}
        \caption{0-dimension }
        \label{fig:pd_dim0}
    \end{subfigure}
    \begin{subfigure}[t]{0.32\textwidth}
       \centering
       \includegraphics[width=4cm]{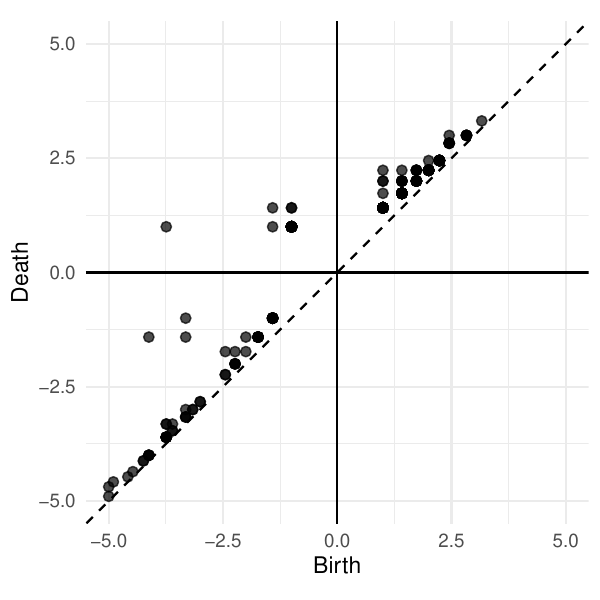}
       \caption{1-dimension }
       \label{fig:pd_dim1}
    \end{subfigure}  
    \begin{subfigure}[t]{0.32\textwidth}
       \centering
       \includegraphics[width=4cm]{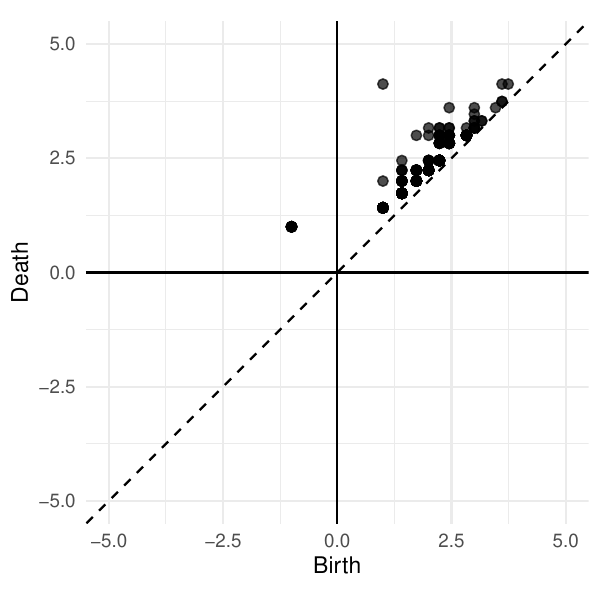}
       \caption{2-dimension }
       \label{fig:pd_dim2}
    \end{subfigure}   
 
  \caption{
    (a)-(j): Visualization of constructing a sublevel set at filtrations $\epsilon\in\{-6,-5,-4,-3,-1,1,2,3,4\}$. While the computation was implemented based on the 3D segmented image, the evolution was described in 2D slices, cut across the red plane in Figure \ref{fig:tumor3d}, for simpler visualization; (k)-(m): persistence diagrams of each dimension summarizing the evolution of topological features in the example tumor image.}
\label{fig:filt}
\end{figure}

The filtered cubical complex representation provides the basis for applying persistent homology to characterize the topological features of the image.
The cubical subcomplex $\mathcal{C}_\epsilon$ is a sublevel set of the filtered complex that only includes the elementary cubes whose filtration values are less than or equal to $\epsilon$. 
The nested sequence of subcomplexes $\{\mathcal{C}_{\epsilon}\}_{\epsilon\in\mathbb{R}}$, satisfying $\mathcal{C}_s \subseteq \mathcal{C}_t$ whenever $s \le t$, captures richer and more detailed shape information on 
how the shape of a tumor evolves over filtration, as illustrated in Figures \ref{fig:filt-5}–\ref{fig:filt5}. 

\subsection{Interpretation of topological shape features}

The birth and death values of topological features obtained from persistent homology applied to distance-transformed images encode quantitative information about the geometric scale and structural configuration of the underlying objects. Prior studies \citep{Robins2016, Thakur2021} have shown that analyzing the birth, death, and persistence of these features yields meaningful insights into the microstructural organization of complex materials. In a similar manner, the topological features derived from brain tumor images can be interpreted to characterize morphological variations within tumor subregions. We note that, in our study, the size of a topological feature corresponds to the radius of the largest sphere that can be inscribed within the region, as the filtration is constructed based on the SEDT-3 values.

\begin{figure}[!ht]
  \centering
    \includegraphics[width=1\linewidth]{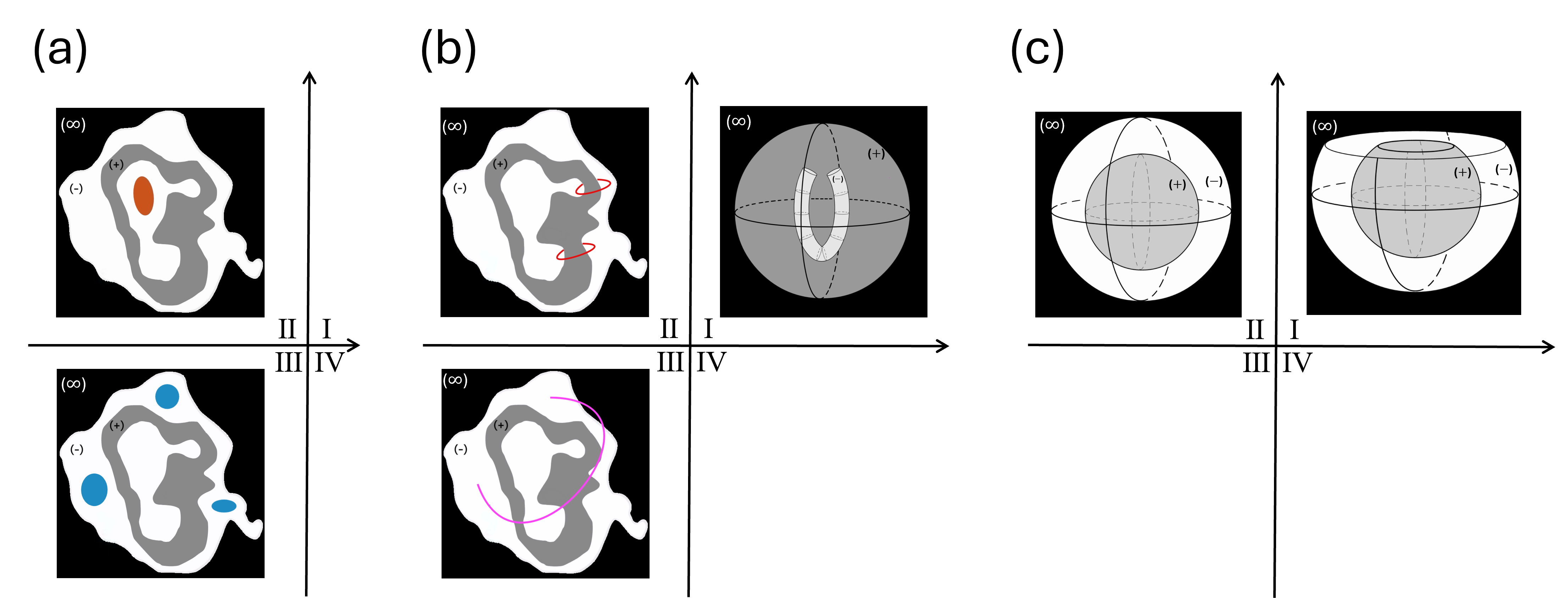}
    \caption{Example patterns that correspond to each quadrant of the persistence diagram for (a) 0-dimensional, (b) 1-dimensional, and (c) 2-dimensional. The non-tumor, non-AT (NCR or ED) and AT regions are colored in black ($\infty$), white (-), and gray (+), respectively. }
    \label{fig:interpretation}
\end{figure}


The extracted topological features can be summarized according to their dimensionality and quadrant on the persistence diagram. Figure~\ref{fig:interpretation} illustrates example tumor patterns corresponding to different quadrants of the diagrams. 
First, 0-dimensional features characterize disconnected non-AT regions (Quadrant II) and connected non-AT regions (Quadrant III). 
Their birth and death values reflect the size and degree of connectedness among tumor subregions. 
Second, one-dimensional features capture loop-like structures, including broken or complete ring-shaped formations associated with necrotic cores (Quadrant I), loop of non-AT regions (Quadrant II), or biconcave disc-shaped non-AT region (Quadrant III). These features correspond to interpretable morphological patterns of non-AT and AT regions. 
Lastly, 2-dimensional features describe enclosed or shell-like configurations, where AT regions are completely encapsulated (Quadrant II) or partially encapsulated by non-AT regions (Quadrant I), indicating the overall volume of AT regions. 
Table~\ref{tab:interpretation} summarizes their interpretations. 

\begin{table}[!ht]
\caption{Topological features and their interpretation across dimensions and quadrants of the persistence diagram.}
\label{tab:interpretation}
\centering
\renewcommand{\arraystretch}{0.7} %
\begin{tabular}{ccl}
\hline
Dim                & Quadrant                                                    & \multicolumn{1}{c}{Feature and Interpretation}                                                                                                                                                                                \\ \hline
\multirow{2}{*}{0} & \MakeUppercase{\romannumeral 2} & \begin{tabular}[c]{@{}l@{}}Disconnected AT regions.\\ $\bullet$ birth: size of disconnected non-AT region.\\ $\bullet$ death: size of AT region between disconnected non-AT regions.\end{tabular}                                                     \\ \cline{2-3} 
                   & \MakeUppercase{\romannumeral 3} & \begin{tabular}[c]{@{}l@{}}Connected non-AT regions.\\ $\bullet$ birth: size of connected non-AT regions. \\ $\bullet$ death: size of contact surface area of connected non-AT region.\end{tabular}                        \\ \hline
\multirow{3}{*}{1} & \MakeUppercase{\romannumeral 1} & \begin{tabular}[c]{@{}l@{}}Broken ring-shaped non-AT regions.\\ $\bullet$ birth: size of gap region. \\ $\bullet$ death: inner diameter of the broken ring.\end{tabular}                                                                              \\ \cline{2-3} 
                   & \MakeUppercase{\romannumeral 2} & \begin{tabular}[c]{@{}l@{}}Loop of non-AT region surrounding AT region\\ $\bullet$ birth: thickness of loop of non-AT region.\\ $\bullet$ death: size of enclosed AT.\end{tabular}                                                                    \\ \cline{2-3} 
                   & \MakeUppercase{\romannumeral 3} & \begin{tabular}[c]{@{}l@{}}Biconcave disc-shaped non-AT region.\\ $\bullet$ birth: thickness of peripheral rim of biconcave disc-shaped non-AT region.\\ $\bullet$ death: size of concave center of biconcave disc-shaped non-AT region.\end{tabular} \\ \hline
\multirow{2}{*}{2} & \MakeUppercase{\romannumeral 1} & 
\begin{tabular}[c]{@{}l@{}}AT region partially encapsulated by non-AT region.\\ $\bullet$ birth: size of exposed surface area of AT region \\ $\bullet$ death: size of partially encaptulated AT region.\end{tabular} 
                                                     \\ \cline{2-3} 
                   & \MakeUppercase{\romannumeral 2} &   
                   \begin{tabular}[c]{@{}l@{}}AT-region completely encapsulated by non-AT region.\\ $\bullet$ birth: thinkness of encapsulating non-AT region \\ $\bullet$ death: size of encapsulated AT region.\end{tabular} \\ \hline
\end{tabular}
\end{table}

\subsection{Functional representation of topological shape features}
\label{subsec:ps}

Although persistence diagrams encode topological features that provide an interpretable representation of tumor morphology, they are abstract algebraic objects rather than conventional data types commonly used in statistical modeling. As a result, they cannot be directly incorporated as covariates in standard statistical or machine learning models. To overcome this limitation, we treat persistence diagrams as functional data by representing them in a functional space, as shown in \cite{chen2015statistical,adams2017persistence}.

Suppose that we have the $j$-dimensional persistent diagram $P^{(j)} =\{(b,d)\in\mathbb{R}^2:b<d\}$, where $b$ and $d$ are filtration values of birth and death, respectively. The persistent surface function $X^{(j)}$ of the persistent diagram $P$ is defined as $X^{(j)}(x,y) = \sum\limits_{(b,d)\in P^{(j)}} g(x,y;b,d,
\sigma_j)\cdot w(b,d)$, where $g(\cdot,\cdot;b,d,\sigma_j)$ is a smoothing kernel for $(b,d)\in P^{(j)}$, and $w(b,d)$ is a non-negative weight function. Throughout this article, we use the Gaussian kernel $g(x,y;b,d,
\sigma_j) = \frac{\exp\left[-(x-b)^2-(y-d)^2\right]}{\sigma_j^2}$ with a smoothing parameter $\sigma_j$ and the maximum distance weight (MDW) function, defined as $w(b,d) = \text{max}\{|b|,|d|,d-b\}$, which assigns higher weights on points farther away from the origin. We refer readers to \citep{Obayashi2018, Moon2023} for investigating other alternative weight functions. To handle features with infinite death values, we replace them with their corresponding birth values, i.e., $(b,d)=(b,\infty)$ is converted to $(b,d)=(b,b)$. This adjustment retains the birth information under the MDW. For each 3D brain tumor images, we have three persistent surface functions; 0-dimension, 1-dimension, and 2-dimension, denoted by $X^{(0)}_i$, $X^{(1)}_i$, and $X^{(2)}_i$, respectively. 

\subsection{Penalized Cox regression with functional predictors}
The Cox regression model has been perhaps the most widely used method to examine effects of covariates on survival outcomes \citep{Cox1972}. In the Cox regression model, the hazard function $h(t)$ has a multiplicative form in relation to $p$-dimensional scalar predictor $\boldsymbol{z}=(z_1,\cdots,z_p)^T$,
\begin{align*}
h(t) = h_0(t)\exp\left(\boldsymbol{z}^{T}\boldsymbol{\alpha}\right),
\end{align*}
$h_0(t)$ is an unspecified baseline hazard function and $\boldsymbol{\alpha}$ is a $p$-dimensional coefficient vector. In references, functional predictors are additionally incorporated into the model besides clinical variables $\boldsymbol{z}$. In particular, following \cite{Moon2023} where persistent surfaces obtained from 2D brain and lung cancers are included as functional predictors, we consider including 0$^{th}$, 1$^{st}$, and $2^{nd}$ persistent surfaces as functional predictors as follows. 
\begin{align}
    h(t)=h_0 (t)\exp\left(\boldsymbol{z}^{T}\boldsymbol{\alpha}+ \sum\limits_{j=0}^{2}\int_{\mathcal{X}^{(j)}} X^{(j)}(u)\beta^{(j)} (u)du\right) \label{eqn:fcox}
\end{align}
Here $\beta^{(j)} (u)$ is the functional coefficient for $j$-th dimension defined over the domain $\mathcal{X}^{(j)}$ for $j=0,1,2$. Hereafter, the argument $u$ of the persistent surface function $X^{(j)}$ is assumed to belong to $\mathcal{X}^{(j)}$ unless stated otherwise.

Because the persistent surfaces $X^{(0)}$, $X^{(1)}$, and $X^{(2)}$ in (\ref{eqn:fcox}) are infinite-dimensional objects, they need to be projected into a finite-dimensional space through functional principal component analysis (FPCA) technique. Let $j$-th dimensional persistent surface function of the $i-$th individual, denoted by $X^{(j)}_i$, have the mean function $\mu^{(j)} (u) = \mathbb{E}\left[X^{(j)}_i (u)\right]$, and spectral decomposition of the covariance function $\text{Cov}\left(X^{(j)}_i(u),X^{(j)}_i (u')\right) = \sum\limits_{k=1}^{\infty}\lambda^{(j)}_{k}\phi^{(j)}_k (u)\phi^{(j)}_k (u')$, where $\{\lambda^{(j)}\}_{k\geq 1}$ is a non-decreasing sequence of non-negative eigenvalues and $\{\phi^{(j)}\}_{k\geq 1}$ is a set of corresponding Eigen functions. By applying the Karhunen-Lo\'eve expansion, we can rewrite the persistent function as $X^{(j)}_{i}(u) = \mu^{(j)}(u)+\sum\limits_{k=1}^{\infty}\epsilon^{(j)}_{ik}\phi^{(j)}_{k}(u)$, where $\epsilon^{(j)}_{ik} = \int_{\mathcal{X}^{(j)}}\left( X^{(j)}_{i}(u)-\mu^{(j)}(u) \right) \phi^{(j)}_{k} (u) du$ is the FPC score of the $j$-th dimension. Note that FPCs have zero mean and orthogonal covariance function, satisfying $\mathbb{E}\left[\epsilon^{(j)}_{ik}\right]=0$ and $\mathbb{E}\left[\epsilon^{(j)}_{ik}\epsilon^{(j)}_{ik'}\right]=\lambda^{(j)}_k \mathbb{1}\left(
k=k'\right)$. Truncating the expansion at a finite order enables an approximation of the persistent surface function $X^{(j)}_i (u)$ of the infinite dimension as a finite function, $X^{(j)}_i (u) = \mu^{(j)}(u) + \sum\limits_{k=1}^{r_j}\epsilon^{(j)}_{ik}\phi^{(j)}_{k}$, where $r_j$'s are the numbers of FPCs selected by a pre-specified threshold. Then, by plugging these approximated functional predictors, the FCox regression model (\ref{eqn:fcox}) can be rewritten as 
\begin{align}
   h(t) = h^{*}_0(t)\exp\left(\boldsymbol{z}^{T}\boldsymbol{\alpha} + \sum\limits_{j=0}^{2}\sum\limits_{k=1}^{r_j} \epsilon^{(j)}_{ik}\beta^{(j)}_{k}\right), \label{eqn:fcox_approx}
\end{align}
where $h^{*}_{0}(t) = h_{0}(t)\exp\left(\sum_{j=0}^{2} \int_{\mathcal{X}^{(j)}} \mu^{(j)}(u)\beta^{(j)} (u)du \right)$. Unlike the infinite dimensional predictors in (\ref{eqn:fcox}), predictors in (\ref{eqn:fcox_approx}) are of the finite dimension, $p+\sum\limits_{j=0}^2 r_j$. When the proportion of variance explained by the first $r_j$ FPCs is defined as $PV^{(j)}(r)=\sum\limits_{k=1}^{r}\lambda^{(j)}_k / \sum\limits_{k=1}^{\infty}\lambda^{(j)}_k$ for the $j$-th dimension, $r_j$ is determined as the minimum order at which $PV^{(j)}$ is larger than a given threshold $C$, i.e., $r_j = \text{min}\{r\text{: } 
PV^{(j)}(r)>C\}$. Throughout this article, $C=90\%$ is assumed. Once the number of FPCs are determined, the partial likelihood for (\ref{eqn:fcox_approx}) can be constructed as follows.
\begin{align} 
 \ell\left(\boldsymbol{\alpha},\boldsymbol{\beta}\right)&= \ell\left(\alpha_1,\cdots,\alpha_p, \beta^{(0)}_1,\cdots, \beta^{(0)}_{r_0}, \beta^{(1)}_{0},\cdots, \beta^{(1)}_{r_1}, \beta^{(2)}_{1},\cdots, \beta^{(2)}_{r_2};r_0,r_1,r_2\right) \notag \\ 
 &=\prod_{i=1}^{n}\frac{ \exp \Big( \boldsymbol{z}_i^{T}\boldsymbol{\alpha} + \sum\limits_{j=0}^{2}\sum\limits_{k=1}^{r_j}\epsilon^{(j)}_{ij}\beta^{(j)}_k \Big)  }{ \sum\limits_{i' \in R_{t_i}}\exp\Big( \boldsymbol{z}_{i'}^{T}\boldsymbol{\alpha} + \sum\limits_{j=0}^{2}\sum\limits_{k=1}^{r_j}\epsilon^{(j)}_{i'j}\beta^{(j)}_k \Big) },\label{eqn:partialLik}
\end{align}
where $t_i$ is $i$-th patient's survival time and $R_{t_i}$ means the risk sets at the time point $t_i$.


Even after the low-dimensional summary of persistent surfaces is obtained through FPCA, the truncated number of FPCs can still pose difficulty in parameter estimation due to the small sample size. It is also expected that, among the selected modes of variation in FPCA, just a few would be meaningfully associated with survival outcomes, which further justifies the introduction of sparsity. Thus, we choose to use the penalized Cox regression based on the $\ell_1$-penalized partial likelihood. 
\begin{align} 
 \ell_p\left(\boldsymbol{\alpha},\boldsymbol{\beta};\lambda\right)&= \ell_p\left(\alpha_1,\cdots,\alpha_p, \beta^{(0)}_1,\cdots, \beta^{(0)}_{r_0}, \beta^{(1)}_{0},\cdots, \beta^{(1)}_{r_1}, \beta^{(2)}_{1},\cdots, \beta^{(2)}_{r_2};r_0,r_1,r_2,\lambda\right) \notag \\ 
 &=\prod_{i=1}^{n}\frac{ \exp \Big( \boldsymbol{z}_i^{T}\boldsymbol{\alpha} + \sum\limits_{j=0}^{2}\sum\limits_{k=1}^{r_j}\epsilon^{(j)}_{ij}\beta^{(j)}_k \Big)  }{ \sum\limits_{i' \in R_{t_i}}\exp\Big( \boldsymbol{z}_{i'}^{T}\boldsymbol{\alpha} + \sum\limits_{j=0}^{2}\sum\limits_{k=1}^{r_j}\epsilon^{(j)}_{i'j}\beta^{(j)}_k \Big) } + \lambda\sum\limits_{j=0}^{2}\sum\limits_{k=1}^{r_j}\big|\beta^{(j)}_{k}\big|, \label{eqn:pp_Lik}
\end{align}
The theory and implementation of penalized Cox regression are well presented in \cite{Simon2011}. The functional coefficient estimate $\widehat{\beta}^{(j)}(u)$ can be computed as $\widehat{\beta}^{(j)}(u)\approx\sum\limits_{k=1}^{r_j}\widehat{\beta}^{(j)}_{k}\widehat{\phi}^{(j)}_k(u)$.


\subsection{Lobe-specific modeling}
\label{subsec:lobe_specific}
Most gliomas are found in the cerebral hemispheres that comprise the frontal, temporal, parietal, and occipital lobes in adults, and are notably less common in the cerebellum \citep{larjavaara2007incidence, tamimi2017, grochans2022}. 
As each lobe supports different brain functions, some studies \citep{Fyllingen2021, YERSAL2017123,simpson1993influence,bao2023new} have investigated the potential prognostic value of tumor involvement in each lobe. In particular, a growing body of evidence \citep{PALDOR2016} suggests that frontal lobe involvement is not only meaningfully associated with survival outcomes, but also may exhibit distinct biological characteristics different from those in other lobes. 

To further investigate, we include such spatial information into the model by adding an indicator function $\mathbb{1}(\texttt{frontal}_i)$ that takes 1 if the $i$-th subject's tumor is located in the frontal lobe and 0 otherwise. Here, a tumor is regarded as belonging to the frontal lobe if the largest portion of the enriching and necrotic areas is spanned over the frontal lobe. Moreover, including an interaction term between the membership indicator $\mathbb{1}(\texttt{frontal}_i)$ and topological features may allow for location-specific effects of topological shape features, leading to the following model. 
\begin{align}
    h_i(t)&=h_0 (t)\exp\left(\boldsymbol{z}_i^{T}\boldsymbol{\alpha}+ \int_{\mathcal{X}^{(j)}} X_{i}^{(j)}(u)\big(\beta^{(j)}(u)+\mathbb{1}(\texttt{frontal}_i)\cdot\beta^{(j)}_{f}(u)\big)du\right)  \notag \\
    &\approx h^{*}_0(t)\exp\left(\boldsymbol{z}_i^{T}\boldsymbol{\alpha} + \sum\limits_{j=0}^{2}\sum\limits_{k=1}^{r_j} \epsilon^{(j)}_{ik}\left(\beta^{(j)}_{k}+\mathbb{1}(\texttt{frontal}_i )\cdot \beta^{(j)}_{k,f} \right)\right), \label{eqn:fcox_approx_lobe}
\end{align}
where $\mathbb{1}(\texttt{frontal}_i)$ is newly included in the predictor vector $\mathbf{z}$. The second line comes from applying the same FPCA technique used in (\ref{eqn:fcox_approx}). 
Then, the $\ell_1$-penalized partial likelihood for (\ref{eqn:fcox_approx_lobe}) is given as
{\footnotesize
\begin{align*} 
\ell_p\left(\boldsymbol{\alpha},\boldsymbol{\beta};\lambda\right)&= \ell_p\left(\alpha_1,\cdots,\alpha_p, \beta^{(0)}_1,\cdots, \beta^{(0)}_{r_0}, \beta^{(1)}_{0},\cdots, \beta^{(1)}_{r_1}, \beta^{(2)}_{1},\cdots, \beta^{(2)}_{r_2};r_0,r_1,r_2,\lambda\right)  \\ \notag
 &=\prod_{i=1}^{n}\frac{ \exp\left\{ \boldsymbol{z}_i^{T}\boldsymbol{\alpha} + \sum\limits_{j=0}^{2}\sum\limits_{k=1}^{r_j}\epsilon^{(j)}_{ij}\left(\beta^{(j)}_k + \mathbb{1}(\texttt{frontal}_i)\cdot \beta^{(j)}_{k,f}\right) \right\}  }{ \sum\limits_{i' \in R_{t_i}}\exp\left\{ \boldsymbol{z}_{i'}^{T}\boldsymbol{\alpha} + \sum\limits_{j=0}^{2}\sum\limits_{k=1}^{r_j}\epsilon^{(j)}_{i'j}\left(\beta^{(j)}_k + \mathbb{1}(\texttt{frontal}_{i'})\cdot \beta^{(j)}_{k,f}\right) \right\} } + \lambda\sum\limits_{j=0}^{2}\sum\limits_{k=1}^{r_j}\left(\big|\beta^{(j)}_{k}\big|+\big|\beta^{(j)}_{k,f}\big|\right),
\end{align*}}
where $t_i$ is $i$-th patient's survival time and $R_{t_i}$ means the risk sets at the time point $t_i$. The partial ML estimates of functional coefficients can be obtained as $\widehat{\beta}^{(j)}(u)\approx\sum\limits_{k=1}^{r_j}\widehat{\beta}^{(j)}_{k}\widehat{\phi}^{(j)}_k(u)$ if a tumor lies in non-frontal lobe regions, and $\widehat{\beta}^{(j)}(u)+\widehat{\beta}^{(j)}_f(u)\approx\sum\limits_{k=1}^{r_j}\left(\widehat{\beta}^{(j)}_{k}+\mathbb{1}(\texttt{frontal}_i)\cdot \widehat{\beta}^{(j)}_{k,f}\right)\widehat{\phi}^{(j)}_k(u)$ if a tumor belongs to the frontal lobe.

\subsection{Selection of hyper-parameters}
\label{subsec:hyperpars}
The Gaussian smoothing parameters $(\sigma_0, \sigma_1, \sigma_2)$ and the $\ell_1$-penalty parameter $\lambda$ are optimized with respect to predictive performance. First, $\lambda$ is profiled over the grid of $(\sigma_0, \sigma_1, \sigma_2)$ by minimizing the 10-fold cross-validated partial log-likelihood deviance \citep{Simon2011}. The grid of $(\sigma_0, \sigma_1, \sigma_2)$ ranges from 0.3 to 3 with a step size of 0.3. Then, a grid search is performed for $(\sigma_0, \sigma_1, \sigma_2)$. Specifically, relative predictive risks are estimated via leave-one-out cross-validation (LOOCV), and subjects are classified into distinct high- and low-risk groups at the median risk value, followed by a log-rank test comparing the two groups. The set of smoothing parameters that yields the smallest p-value in the log-rank test is selected as optimal.

\section{Simulation study}
\label{sec:simul}

\begin{figure}[!ht]
    \centering
    \begin{subfigure}[t]{0.3\textwidth}
        \centering
        \includegraphics[width=6cm]{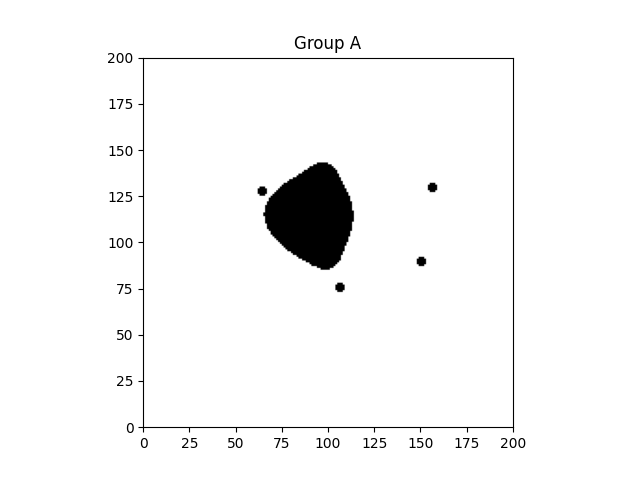}
        \caption{Group A}
        \label{fig:example_A}
    \end{subfigure}
        \begin{subfigure}[t]{0.3\textwidth}
        \centering
        \includegraphics[width=6cm]{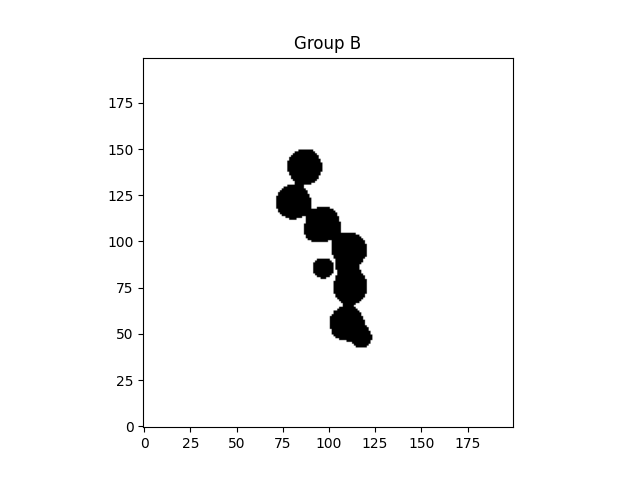}
        \caption{Group B}
        \label{fig:example_B}
    \end{subfigure}
    \caption{Simulated tumor examples for Groups A and B.}
    \label{fig:Example_tumor}
\end{figure}

We conduct a simulation study to assess the performance of the proposed PH-FCox model in capturing shape differences and their association with survival outcomes.
Under the simulation setting, each subject is assumed to belong to one of two distinct brain tumor types, namely, Group A and Group B. Tumors in Group A are characterized by a large central mass accompanied by several satellite lesions, whereas tumors in Group B comprise multiple smaller masses forming an elongated pattern. Example tumor images for each group are provided in Figure~\ref{fig:Example_tumor}.

For each dataset, we set the sample size to 140 with the censoring rate of 15$\%$ to reflect the number of patients in the applications described in Section~\ref{subsec:GBM}. A half of the subjects are assigned to Group A and the remaining half to Group B. Within each group, 30$\%$ of the subjects have tumors located in the frontal lobe. For simplicity, 2-dimensional binary images (tumor vs. non-tumor) of size $200\times200$ pixels are used. A total of 300 datasets are generated.


To create the main tumor body for Group A, we applied bivariate smoothing to 30 points randomly drawn from $N_2\left(\boldsymbol{0}_2\, ,\, 1.25^2 \cdot I_2\right)$, where $\boldsymbol{0}_2$ and $I_2$ denote the 2-dimensional zero vector and identity matrix, respectively. A bivariate Gaussian kernel is used with smoothing parameters $\sigma_x = \sigma_y = 7$. Similarly, small tumor debris surrounding the main body is generated by drawing random points from $N_2\left(\boldsymbol{0}_2\, , \, 5^2 \cdot I_2\right)$, followed by Gaussian smoothing with $\sigma_x = \sigma_y = 2$. Finally, we threshold the resulting density values at 0.0025 for the main body and 0.02 for the small debris.


For Group B, six mean vectors ($\boldsymbol{\mu}_1,\ldots,\boldsymbol{\mu}_6$) are randomly selected such that they are approximately aligned along a straight line in the 2D plane. Then, 2000 random points are generated from each distribution $N_2\left(\boldsymbol{\mu}_i\, , \, 0.4^2 \cdot I_2\right)$ ($i=1,\ldots,6$), followed by Gaussian smoothing with $\sigma_x = \sigma_y = 2$. Smaller tumor fragments surrounding the main body are then simulated in a similar manner by drawing 500 random points from each of $N_2\left(\boldsymbol{\mu}_i\, , \, 4.8^2 \cdot I_2\right)$. Finally, both the main tumor body and fragments are thresholded at a density value of 1. Compared to Group A, a larger number of points is generated, and a higher threshold is used to create a cluster of closely connected small tumor bodies. Tumor sizes between the two groups are set to be comparable, with the mean size of around 2,000 pixels, in order to avoid potential confounding effects.

The survival time for the $i$-th subject ($i=1,\cdots,140$) is generated by assuming an exponential distribution with the following proportional hazard function,
\begin{align}
   h_i(t) = h_0\cdot \exp\big( 0.8\times\mathbb{1}_i(\text{B}, \texttt{frontal}) + 0.4\times\mathbb{1}_i(\text{B}, \texttt{non-frontal}) - 0.2\times\mathbb{1}_i(\text{A},\texttt{frontal}) \big), \label{eqn:surv_time}
\end{align}
where $\mathbb{1}_i(\text{Group}, \text{Location})$ is an indicator function that equals 1 if the $i$-th subject belongs to the specified group and has a tumor located in the specified brain region, and 0 otherwise. For example, $\mathbb{1}_i(B, \texttt{frontal})$ equals 1 if the $i$-th subject belongs to Group B and has a tumor in the frontal lobe. No clinical variables are considered here for simplicity. Under this setup, the hazard ratio of B to A is $e^{1}$ in the frontal lobe and $e^{0.4}$ in the non-frontal lobes, indicating varying effects of tumor shape by location. Following the probability integral transform for proportional hazard models in \cite{Bender2005}, survival times are generated from an exponential distribution with a rate parameter of 1/3000. Censoring times are then independently generated from an exponential distribution such that the censoring rate is 15\%.







In place of the shape groups defined in the true model (\ref{eqn:surv_time}), the PH-FCox model has the functional shape predictors $X^{(0)}(u)$, $X^{(1)}(u)$, and $X^{(2)}(u)$, as follows.
\begin{align}
    h_i(t)&=h_0 (t)\exp\left( \int_{\mathcal{X}^{(j)}} X_{i}^{(j)}(u)\big(\beta^{(j)}(u)+\mathbb{1}(\texttt{frontal}_i)\cdot\beta^{(j)}_{f}(u)\big)du\right)  \notag \\
    &\approx h^{*}_0(t)\exp\left(\sum\limits_{j=0}^{2}\sum\limits_{k=1}^{r_j} \epsilon^{(j)}_{ik}\left(\beta^{(j)}_{k}+\mathbb{1}(\texttt{frontal}_i )\cdot \beta^{(j)}_{k,f} \right)\right). \notag
\end{align}
We expect that these shape features carry information that can capture the difference between the true shape groups and their association with survival risks.
Because simulated images are binary, we use the SEDT-2 \citep{Moon2023} that assigns negative (positive) signs to tumor (non-tumor) pixels with smoothing parameters fixed at $\sigma_0=\sigma_1=2$. The numbers of the FPC scores are selected using the threshold of $C=90\%$ in FPCA.  

\begin{figure}[!ht]
    \centering
    \begin{subfigure}[t]{0.24\textwidth}
        \centering
        \includegraphics[width=4.2cm]{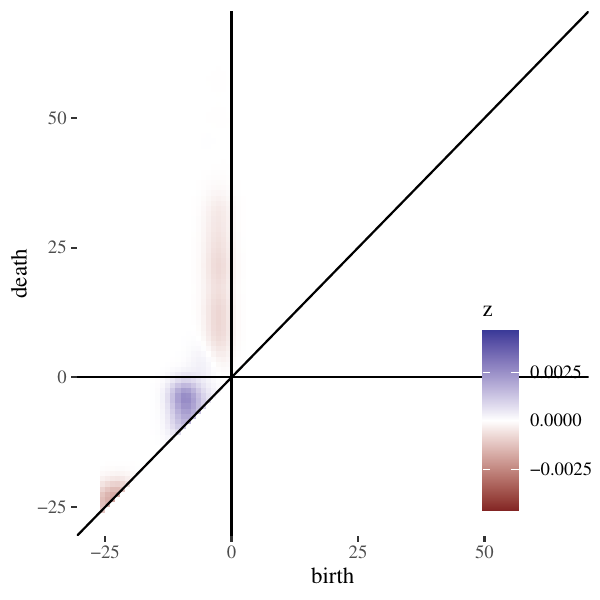}
        \caption{$\widehat{\beta}^{(0)}(u)$}
        \label{fig:simul1}
    \end{subfigure}
    \begin{subfigure}[t]{0.24\textwidth}
        \centering
        \includegraphics[width=4.2cm]{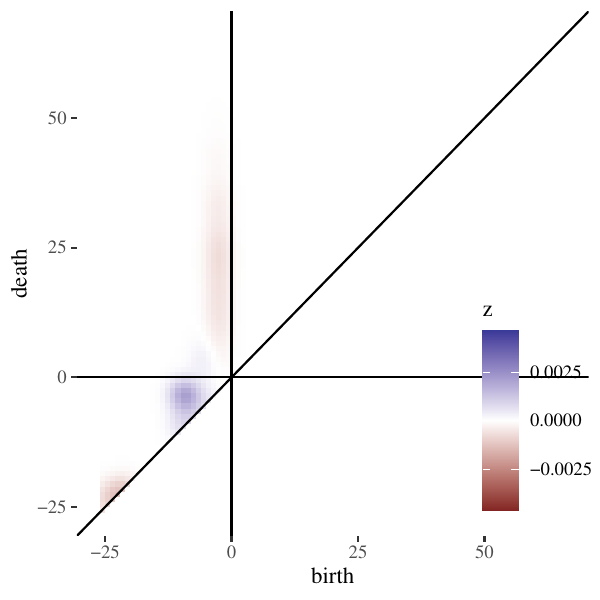}
        \caption{$\widehat{\beta}_{f}^{(0)}(u)$}
        \label{fig:simul2}
        \end{subfigure}
     \begin{subfigure}[t]{0.24\textwidth}
       \centering
       \includegraphics[width=4.2cm]{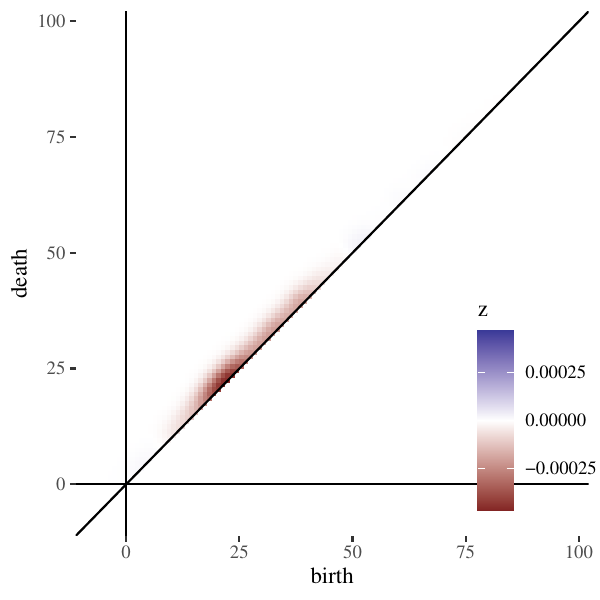}
        \caption{$\widehat{\beta}^{(1)}(u)$}
    \label{fig:simul3}
    \end{subfigure}
     \begin{subfigure}[t]{0.24\textwidth}
       \centering
       \includegraphics[width=4.2cm]{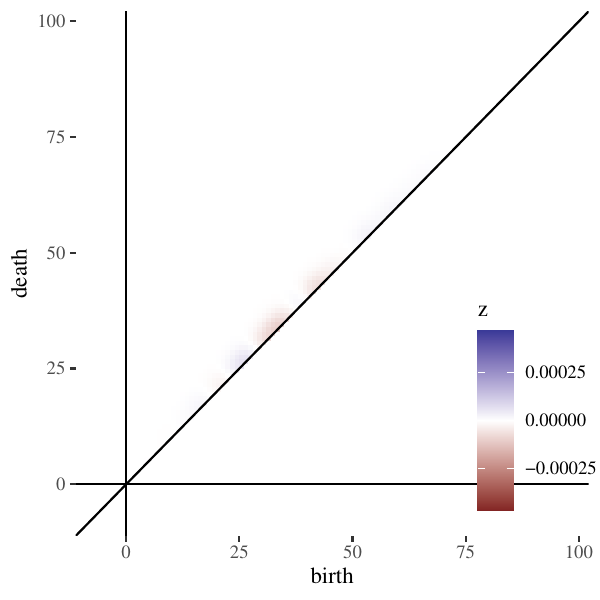}
        \caption{$\widehat{\beta}_{f}^{(1)}(u)$}
    \label{fig:simul4}
    \end{subfigure}
    
    \caption{Functional coefficient estimates averaged over 300 repetitions.}
    \label{fig:simul_fcoef}
\end{figure}

We present the average functional coefficient estimates from 300 simulation repetitions in Figure \ref{fig:simul_fcoef}. The results show that the PH-FCox model can account for 1) the difference in tumor shapes between the two groups and their association with survival risks and 2) varying effects by tumor location. Figure~\ref{fig:simul1} presents the estimated functional coefficients for a 0-dimensional shape feature in the non-frontal lobes. The red areas around birth values of -25 and -2 correspond to connected components formed by the main body and small scattered debris of tumors in Group A, respectively. The blue region between the two red regions represents the connected components formed by a cluster of smaller lesions in Group B. This indicates that Group B is associated with a higher risk of death than Group A, consistent with the true model \eqref{eqn:surv_time}. Also, the functional coefficient estimate for the 0-dimensional interaction term in Figure~\ref{fig:simul2}, denoted by $\widehat{\beta}_{f}^{(0)}(u)$, makes stronger effect sizes in the frontal lobe, summing up to $\widehat{\beta}^{(0)}(u)+\widehat{\beta}_{f}^{(0)}(u)$. With respect to the 1-dimensional result provided in Figure \ref{fig:simul3}, the red regions likely correspond to Group A, where ring-shaped patterns emerge as the main tumor body connects with nearby debris over the filtration. It implies that Group A is associated with a lower risk of death, which aligns with \eqref{eqn:surv_time}.

\section{Clinical Data Analysis}
\label{sec:analysis}

\begin{figure}[!ht]
\centering
        \includegraphics[width=10cm]{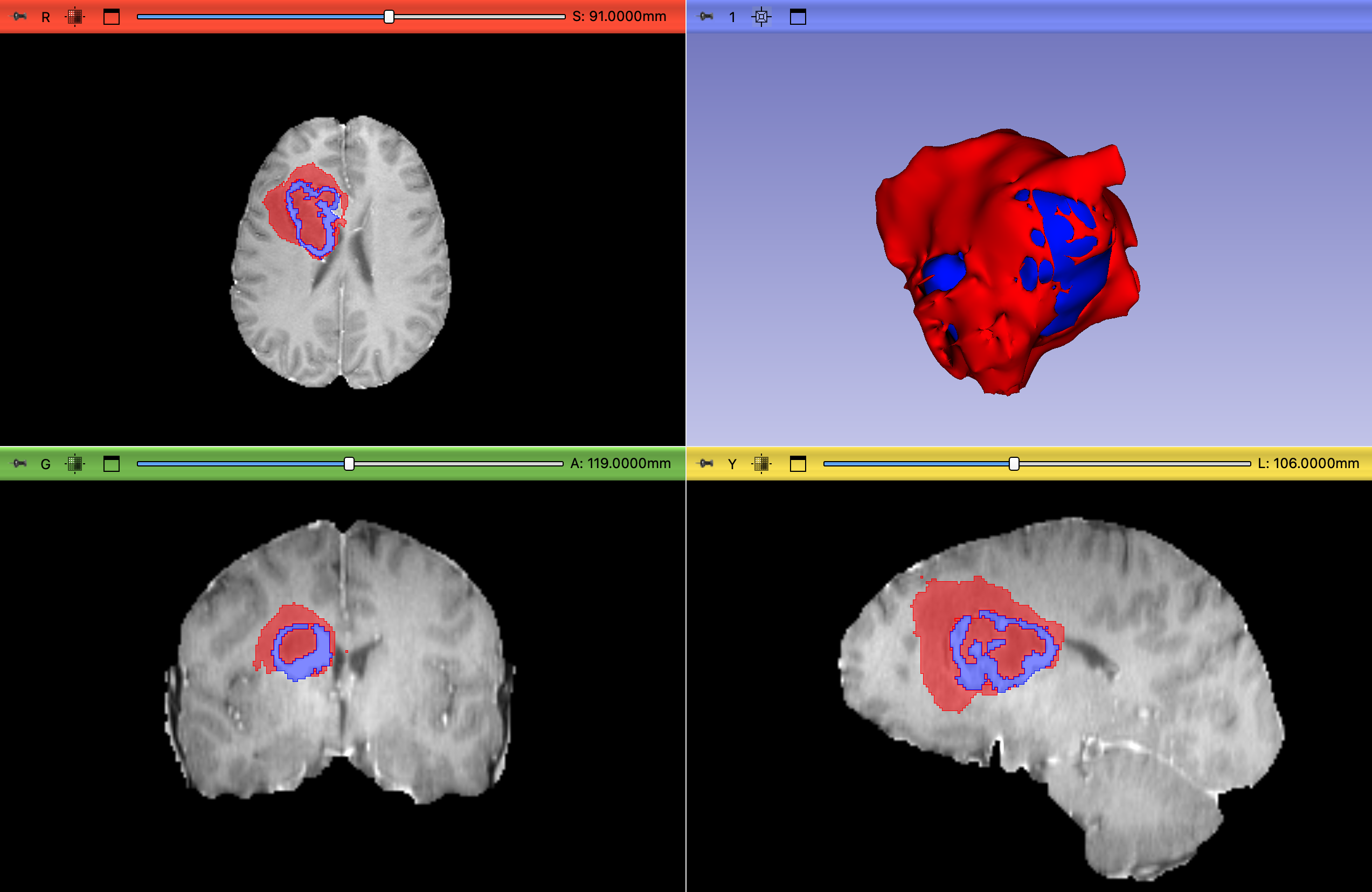}
    \caption{Axial slices of a brain with GBM and its 3D segmented tumor image.}
    \label{fig:tumor_example}
\end{figure}

We use the Brain Tumor Segmentation (BraTS) dataset \citep{bakas2017advancing, bakas2018identifying, menze2014multimodal}, which is publicly available from The Cancer Imaging Archive (TCIA) \citep{Scarpace2016}. The BraTS images provide skull-stripped structural MRI scans with an isotropic resolution of $1 \mathrm{mm}^3$. Tumor segmentation is based on four MRI modalities: T1, T1-weighted, T2, and T2-FLAIR. The multimodal scans were first segmented using the GLISTRboost method \citep{bakas2015glistrboost} and then manually inspected and corrected by experienced neuroradiologists. Figure~\ref{fig:tumor_example} shows an example BraTS image with the 3D segmented tumor and its axial slices. The annotated tumor subregions consist of AT, ED, and NCR/NET, as explained in Section~\ref{subsec:PH}. Because these MRI scans are aligned only in position and orientation by default, we further standardize them to correct for differences in size and shape as outlined in Section~S.2 in the Supplementary Material.

We incorporate four clinical variables, denoted by $\boldsymbol{z}$ in (\ref{eqn:fcox_approx_lobe}): sex, age, Karnofsky Performance Scale (KPS), and tumor volume. Tumor volume, measured as the number of AT voxels, is included to control for the potential confounding effect of tumor size. These clinical variables are excluded from penalization to properly adjust for their effects.

\subsection{Application to GBM MRI images}
\label{subsec:GBM}




The segmented GBM MRI images display a heterogeneous tumor composition with distinct yet interrelated sub-regions \citep{bakas2018identifying}. The AT typically forms a rim surrounding the necrotic center in 2D axial slices and represents a smaller but biologically aggressive portion of the tumor. The NCR/NET occupies the central region and varies in size depending on the extent of necrosis, often becoming substantial in advanced disease. The ED constitutes the largest and most diffuse component, extending irregularly beyond the enhancing margins and reflecting the highly infiltrative nature of GBM.
We conduct the analysis on 133 GBM cases from the BraTS dataset after excluding two subjects due to poor image quality.



\begin{figure}[!ht]
    \centering
    
    \begin{subfigure}[t]{0.24\textwidth}
        \centering
        \includegraphics[width=4.2cm]{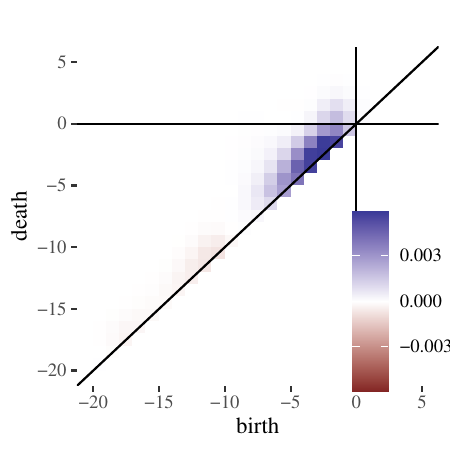}
        \caption{$\widehat{\beta}^{(0)}(u)$}
        \label{fig:gbm_fcoef0}
    \end{subfigure}
    \begin{subfigure}[t]{0.24\textwidth}
        \centering
        \includegraphics[width=4.2cm]{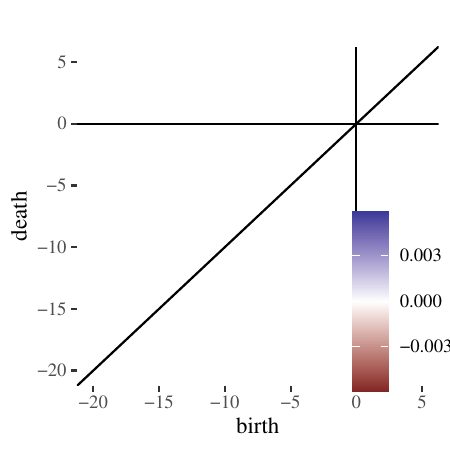}
        \caption{$\widehat{\beta}^{(0)}_{f}(u)$}
        \label{fig:gbm_fcoef0f}
    \end{subfigure}
    \begin{subfigure}[t]{0.24\textwidth}
        \centering
        \includegraphics[width=4.2cm]{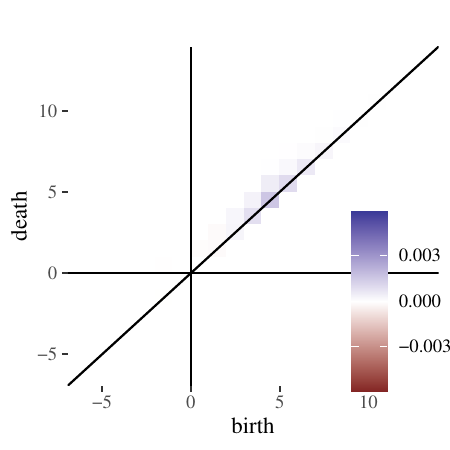}
        \caption{ $\widehat{\beta}^{(2)}(v)$}
        \label{fig:gbm_fcoef2}
    \end{subfigure}
    \begin{subfigure}[t]{0.24\textwidth}
        \centering
        \includegraphics[width=4.2cm]{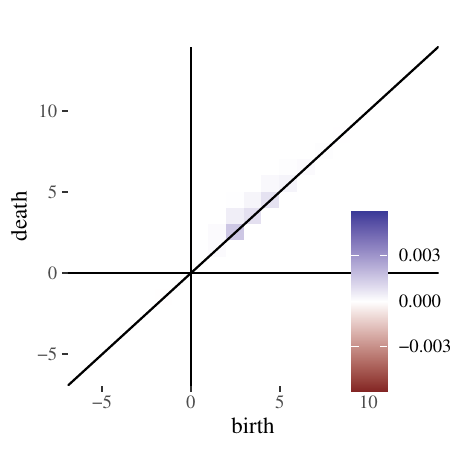}
        \caption{$\widehat{\beta}^{(2)}_{f}(v)$}       
        \label{fig:gbm_fcoef2f}
    \end{subfigure}
    \caption{The 0- and 2-dimensional functional coefficient estimates of the PH-FCox model.}
\label{fig:GBM_coef}
\end{figure}

The PH-FCox model only selects 0- and 2-dimensional topological shape features. The estimated functional coefficients are presented in Figure~\ref{fig:GBM_coef}. It shows how and which 0- and 2-dimensional shape features are associated with the risk of death. Figure~\ref{fig:gbm_fcoef0} shows positive coefficients at birth filtration values between –5 and -1 in Quadrant~\MakeUppercase{\romannumeral 3}, corresponding to connected components of non-AT regions with radii ranging from 1 mm to 5 mm. 
These results suggest that scattered connected components of NCR/NET and irregular ED are generally associated with a higher risk of death. The absence of interaction terms, as shown in Figure~\ref{fig:gbm_fcoef0f}, indicates a lobe-invariant effect of the 0-dimensional feature.

In Figure~\ref{fig:gbm_fcoef2}, the 2-dimensional shape feature exhibits moderately positive coefficients in Quadrant~\MakeUppercase{\romannumeral 1}, which likely correspond to AT partially surrounded by non-AT. Moreover, the death coordinates provide information relevant to the size of AT, as summarized in Table~\ref{tab:interpretation}. The positive interaction effect shown in Figure~\ref{fig:gbm_fcoef2f} suggests a stronger association in frontal lobe tumors. In summary, our findings indicate that a larger number of connected components of AT is associated with a higher risk of death in a lobe-invariant manner, whereas frequent occurrences of non-AT are moderately associated with a higher hazard, particularly in the frontal lobe.

To evaluate the predictive performance of the proposed method, 133 GBM subjects are stratified into high- and low-risk groups based on their predicted death risks. The median predicted risk is used as the threshold, resulting in 67 subjects in the high-risk group and 66 in the low-risk group. 
We assess the predictive efficacy using the p-value of the log-rank test between the high-risk and low-risk groups. 

For comparison, the same procedure is also applied to three existing alternative models: clinical-Cox, radiomic-Cox, and SECT-FCox models.
\begin{itemize}
    \item \textit{clinical-Cox}: The clinical-Cox model is a typical Cox regression with common clinical variables: sex, age, Karnofsky performance scale, and the size of AT. 
    \item \textit{radiomic-Cox}: The radiomic-Cox model incorporates various radiomic features of the TCGA-GBM dataset, such as intensity, volumetric, morphologic, histogram-based, textural, spatial, and glioma diffusion properties \citep{bakas2017advancing}. Out of 726 radiomic variables, 711 were used after excluding those with missing values. We included these radiomic features in addition to the clinical variables and applied a $\ell_1$-penalized Cox regression. The LOOCV for estimating predictive risks is performed by fixing the set of variables selected in the fitted model. 
    \item \textit{SECT-FCox}: The SECT-FCox model implements topological shape features extracted by the Smooth Euler Characteristic Transformation (SECT) \citep{Crawford2020}. The SECT transform encodes topological invariants derived from sub-level sets of the mesh, providing an insightful representation for comparing morphological variations. We computed the SECT of the 3D brain tumor images using the SINATRA proposed by \cite{wang2021statistical}. For a given 3D brain tumor image, the voxels corresponding to the AT, NCR/NET, and ED tumor regions were extracted and represented as three distinct point clouds. We generated three 3D triangular meshes using the point clouds through the marching cubes algorithm \citep{lorensen1998marching}. The SECT curves of the three triangular meshes were computed along multiple uniform directions in cones; we set 15 cones, 4 directions per cone, and 100 sub-levels. The SECT-FCox includes the clinical variables and the FPC scores of the SECT curves, with the number of FPCs truncated at $C=90\%$. The $\ell_1$-penalty term was not applied because more than $95\%$ of the variance was explained by its first FPC. 

\end{itemize}

\begin{figure}[!ht]
    \centering
    
    \begin{subfigure}[t]{0.24\textwidth}
        \centering
        \includegraphics[width=4.2cm]{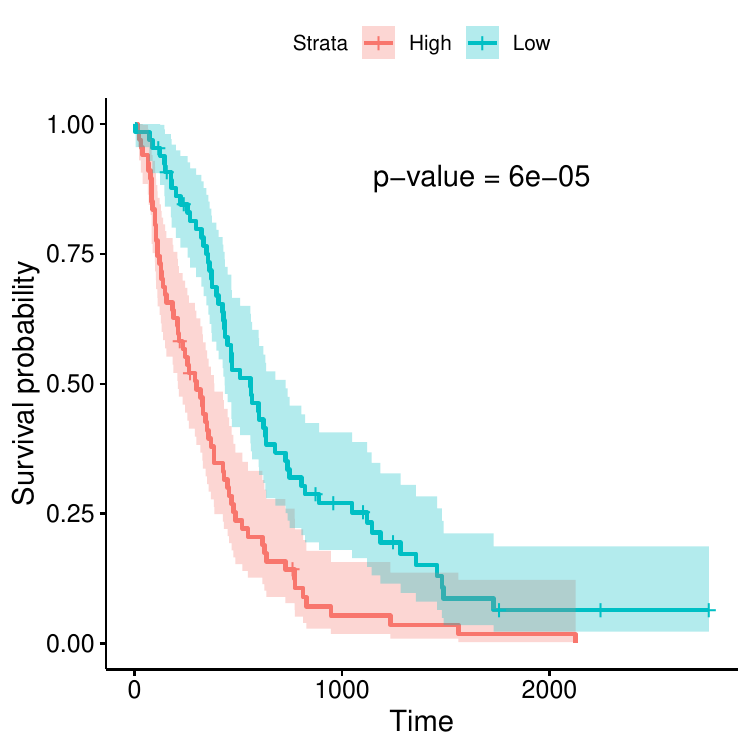}
        \caption{Clinical-Cox}
        \label{fig:cox}
    \end{subfigure}
    \begin{subfigure}[t]{0.24\textwidth}
        \centering
        \includegraphics[width=4.2cm]{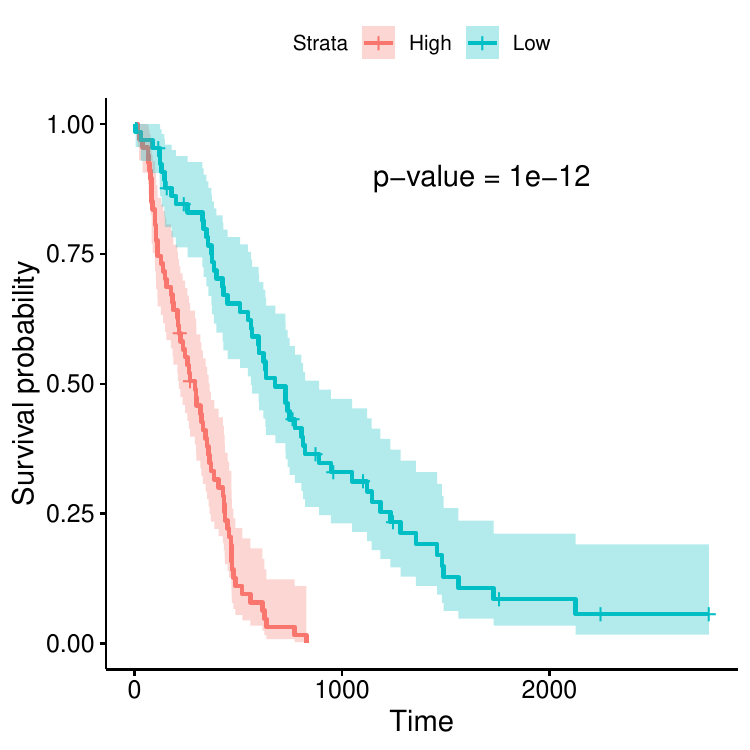}
        \caption{PH-FCox}
        \label{fig:fcox}
    \end{subfigure}
    \begin{subfigure}[t]{0.24\textwidth}
        \centering
        \includegraphics[width=4.2cm]{images/GBM/Surv/GBM_clinical.pdf} 
        \caption{Radiomic-Cox}
        \label{fig:radiomic}
    \end{subfigure}
    \begin{subfigure}[t]{0.24\textwidth}
        \centering
        \includegraphics[width=4.2cm]{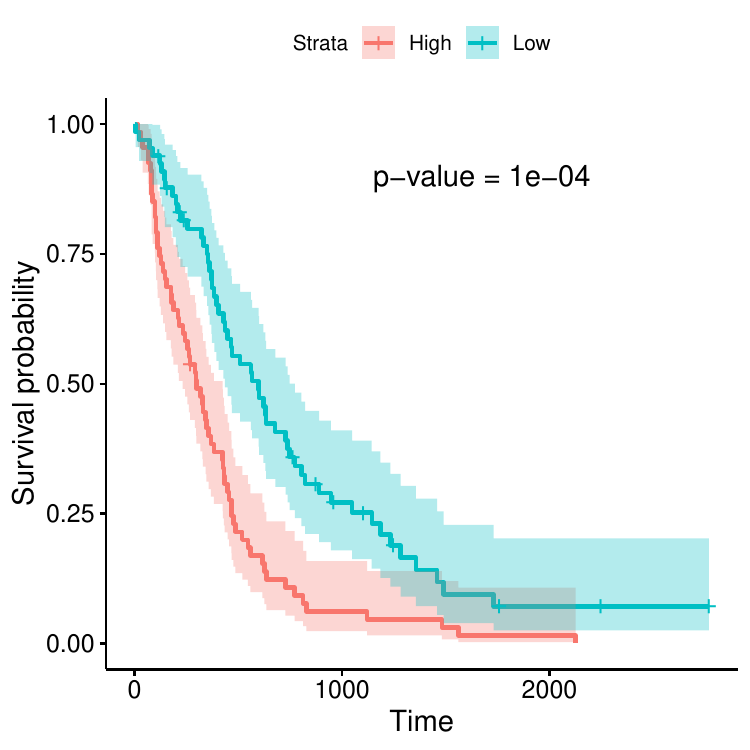}
    \caption{SECT-FCox}
    \label{fig:sect}
    \end{subfigure}
    
    \caption{The Kaplan-Meier plots for the high- and low-risk groups for the GBM data.}
    \label{fig:surv}
\end{figure}

Figure \ref{fig:surv} presents the estimated survival curves for the four models, along with the p-values from the log-rank test. The PH-FCox model performs the best in terms of p-value, followed by the Radiomic-Cox model, the clinical-Cox model, and the SECT-FCox model, in that order. When observing the estimated survival curves, it is obvious that the SEDT-Cox model performs better at the tail. For instance, the confidence intervals in Figure \ref{fig:cox} begin to overlap after 1000 days, whereas Figure \ref{fig:fcox} shows no overlap beyond the early stages of the time axis.

\subsection{Application to LGG MRI images}
\label{subsec:LGG}

We apply the same methodological framework used for GBM to LGG data. Specifically, we analyzed MRI data from 107 LGG subjects, excluding one subject with missing survival information. However, the LGG cohort exhibits several key differences from GBM. First, LGG patients exhibit substantially longer survival, with a censoring rate of approximately 80$\%$, which introduces greater uncertainty. Nevertheless, our model and analytical pipeline are still applicable in the same manner as in the GBM analysis, and it is of particular interest to examine how topological shape features inform the analysis of relatively less complex tumors. Second, the compositional characteristics of LGG differ substantially from those of GBM, reflecting their distinct biological and radiographic profiles. For LGG, segmentation primarily includes NET, as these tumors typically lack NCR and AT regions and show minimal ED. In most cases, the NET region represents the main abnormality, with ED occasionally present, reflecting the limited blood–brain barrier disruption and slow growth of LGG \citep{bakas2018identifying}.
Lastly, KPS was not available for LGG and was therefore excluded from the analysis. 


\begin{figure}[!ht]
    \centering    
    \begin{subfigure}[t]{0.24\textwidth}
        \centering
        \includegraphics[width=4.2cm]{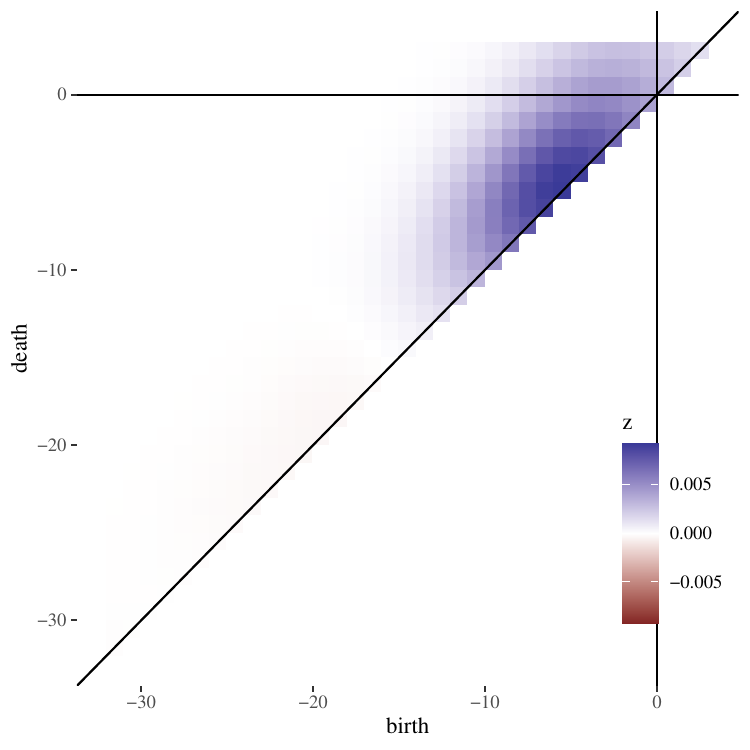}
        \caption{$\widehat{\beta}^{(0)}(u)$}
        \label{fig:lgg_fcoef_dim0}
    \end{subfigure}
    \begin{subfigure}[t]{0.24\textwidth}
        \centering
        \includegraphics[width=4.2cm]{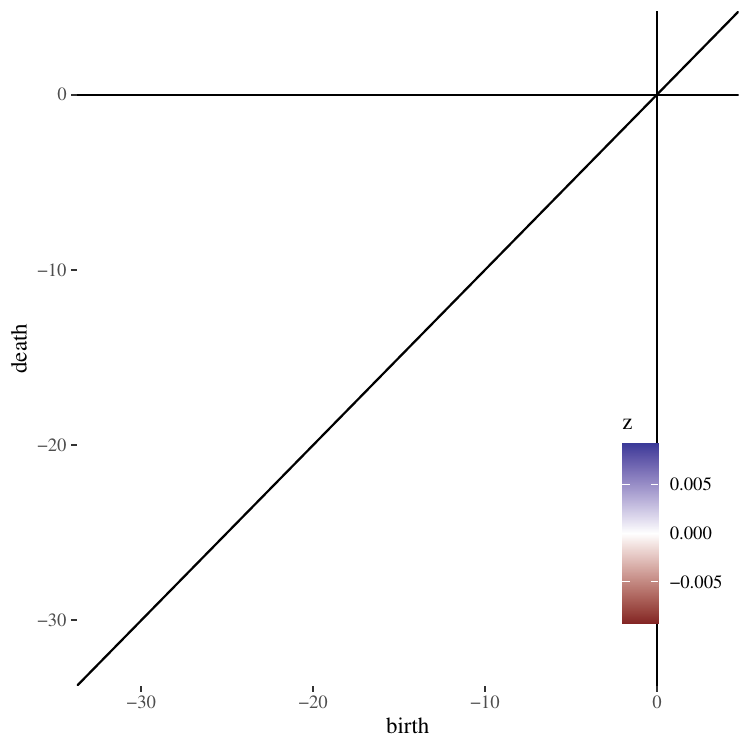}
        \caption{$\widehat{\beta}^{(0)}_{f}(u)$}
        \label{fig:lgg_fcoef_dim0f}
    \end{subfigure}
    \begin{subfigure}[t]{0.24\textwidth}
        \centering
        \includegraphics[width=4.2cm]{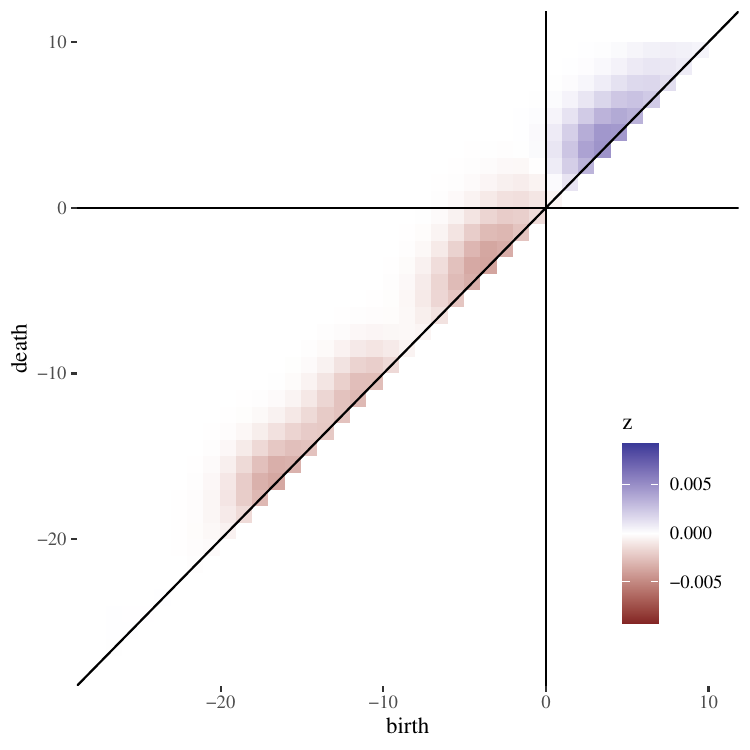}
        \caption{ $\widehat{\beta}^{(1)}(v)$}
        \label{fig:lgg_fcoef_dim1}
    \end{subfigure}
    \begin{subfigure}[t]{0.24\textwidth}
        \centering
        \includegraphics[width=4.2cm]{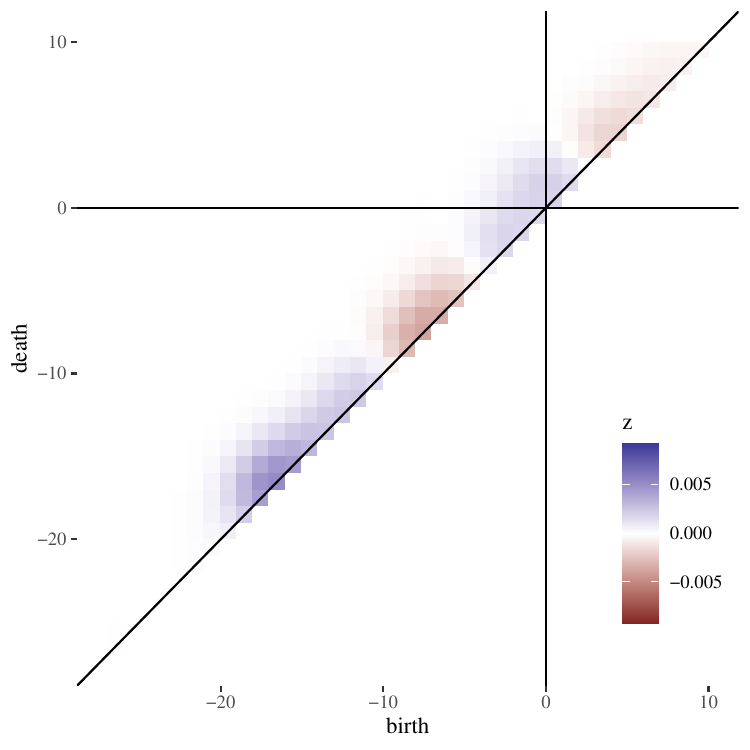}
        \caption{$\widehat{\beta}^{(1)}_{f}(v)$}       
        \label{fig:lgg_fcoef_dim1f}
    \end{subfigure}
    \caption{The 0- and 1-dimensional functional coefficient estimates of the PH-FCox model for the LGG data set.}
\label{fig:LGG_fcoef}
\end{figure}

Figure \ref{fig:LGG_fcoef} shows the estimated functional coefficients obtained from the 107 LGG cohort. 
The most notable difference between GBM is the exclusion of 2-dimensional features and the selection of 1-dimensional features. This outcome is not entirely unexpected, as the 2-dimensional topological features summarized in Table~\ref{tab:interpretation} are primarily driven by moderate volumes of AT region, which is relatively limited in low-grade gliomas. 
The 0-dimensional results presented in Figure \ref{fig:lgg_fcoef_dim0} exhibit distinct positive coefficients at birth values between –10 and 0 in Quadrant \MakeUppercase{\romannumeral 3}. This finding indicates that a larger number of connected components with radii up to 10 mm within the NET region is associated with a higher risk of death, rather than the presence of a single dominant structure. 
Figures~\ref{fig:lgg_fcoef_dim1} and \ref{fig:lgg_fcoef_dim1f} present the estimated functional coefficients of dimension-1 topological features and their interaction with frontal-lobe location. Broken ring-shaped non-AT regions are associated with higher death risk, whereas biconcave-shaped non-AT regions correspond to lower death risk, though the magnitude and regional variation of these effects differ. These findings suggest that morphological irregularity influences prognosis, and that the clinical relevance of structural complexity depends on tumor location.


\begin{figure}[!ht]
    \centering
    
    \begin{subfigure}[t]{0.24\textwidth}
        \centering
        \includegraphics[width=4.2cm]{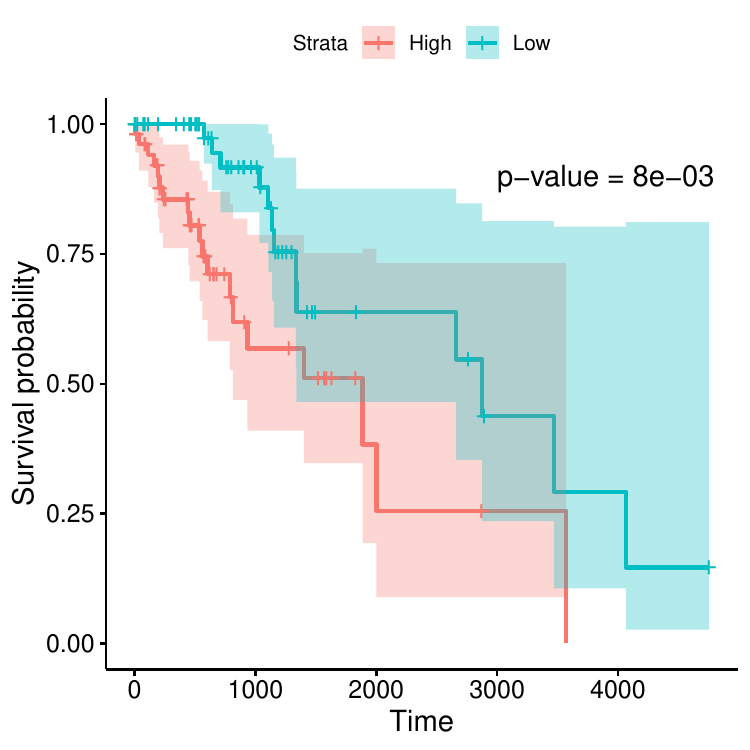}
        \caption{Clinical-Cox}
        \label{fig:lgg_cox}
    \end{subfigure}
    \begin{subfigure}[t]{0.24\textwidth}
        \centering
        \includegraphics[width=4.2cm]{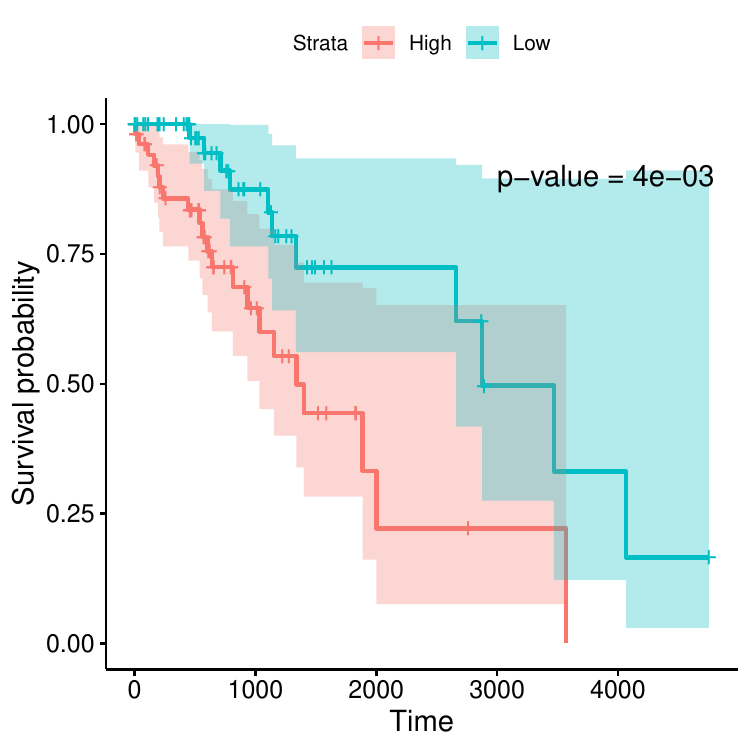}
        \caption{PH-FCox}
        \label{fig:lgg_fcox}
    \end{subfigure}
    \begin{subfigure}[t]{0.24\textwidth}
        \centering
        \includegraphics[width=4.2cm]{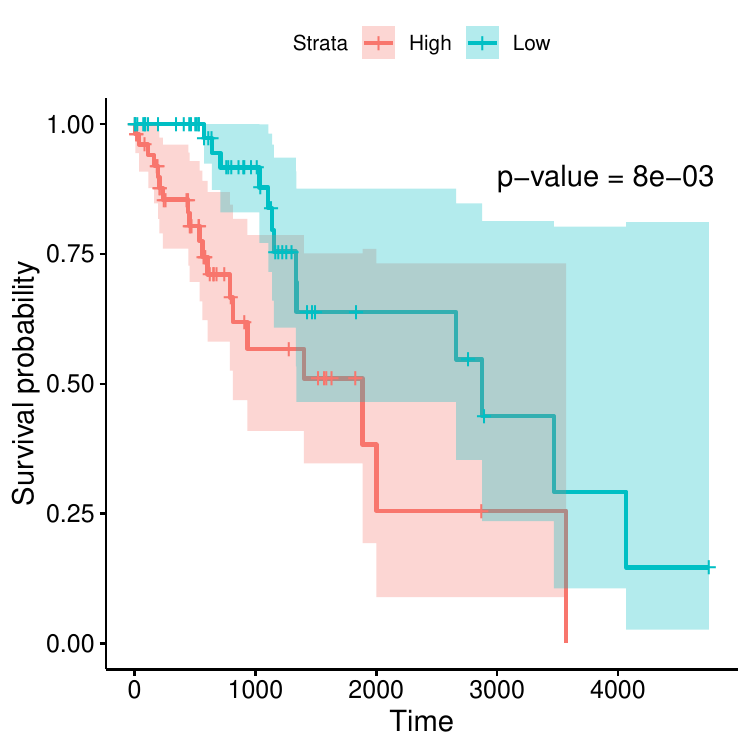}
        \caption{Radiomic-Cox}
        \label{fig:lgg_radiomic}
    \end{subfigure}
    \begin{subfigure}[t]{0.24\textwidth}
        \centering
        \includegraphics[width=4.2cm]{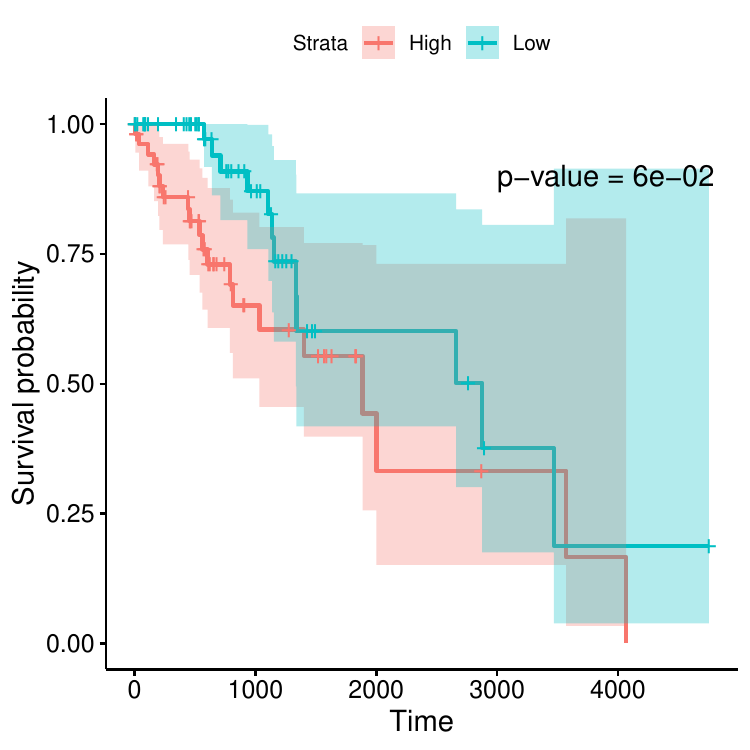}
    \caption{SECT-FCox}
    \label{fig:lgg_sect}
    \end{subfigure}
    
    \caption{The Kaplan-Meier plots for the high- and low-risk groups for the LGG data.}
    \label{fig:LGG_surv}
\end{figure}

The predictive performance of the SECT-FCox model for the LGG dataset is evaluated in comparison with the other alternative models considered in the previous GBM analysis. As summarized in Figure~\ref{fig:LGG_surv}, the PH-FCox model yields the most distinct stratification, as indicated by the smallest $p$-value from the log-rank test. The Radiomic-Cox model is based on a total of 263 radiomic features after discarding those with missing values. As in the GBM analysis, no radiomic feature was selected, resulting in the same $p$-value as the Clinical-Cox regression model. The SECT-FCox model again exhibited the poorest performance, indicating the limited predictive power of SECT.

\subsection{Clinical implications}

We investigated patients whose risk classifications changed under the PH-FCox model compared to their original classification under the clinical-Cox model.
As the PH-FCox model showed better performance for GBM, we focused our analysis on its clinical implications for this cohort.
Specifically, we examined 14 GBM patients reclassified from low to high risk and 14 reclassified from high to low risk when transitioning from the clinical-Cox to the PH-FCox model. We refer to the former as \textit{low-to-high} and the latter as \textit{high-to-low} cases. 

\begin{figure}[!ht]
\centering
    \includegraphics[width=17cm]{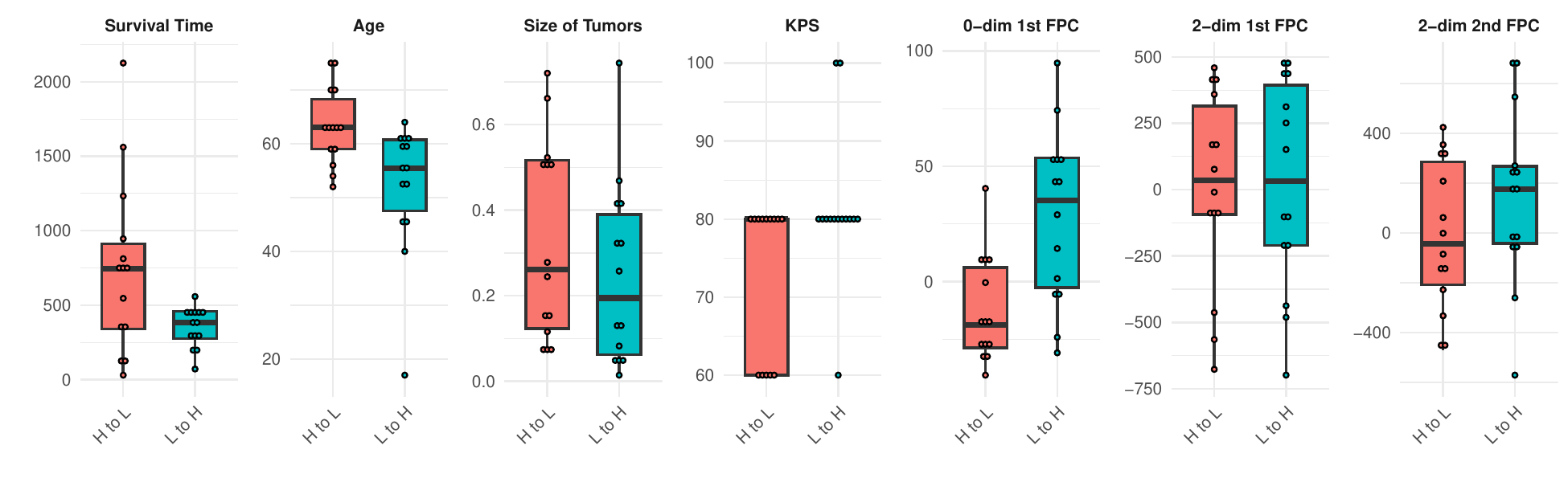}
    \caption{ Box plots of survival times, clinical variables, and FPC scores between the \textit{high-to-low}  and \textit{low-to-high} groups.}
    \label{fig:flip}
\end{figure}

Figure~\ref{fig:flip} compares the survival times, clinical variables, and FPC scores of the two types of flipped cases. The \textit{low-to-high} cases have subjects with younger ages and smaller sizes of tumors on average. However, including topological shape features in the PH-FCox model identifies the high-risk subjects who exhibit relatively shorter survival times. The main difference arises from the 0-dimensional first FPC scores in Figure~\ref{fig:gbm_fcoef0}, which reflect a larger number of non-active tumors' connected components with radii up to 5 mm in the high-risk cases.
Although not as strong as in the 0-dimensional feature, a difference in the 2-dimensional feature between the \textit{high-to-low}  and \textit{low-to-high} groups appears to exist, particularly in 2nd FPC. 


\begin{figure}[!ht]
    \centering
    \begin{subfigure}[t]{0.23\textwidth}
        \centering
        \includegraphics[width=4cm]{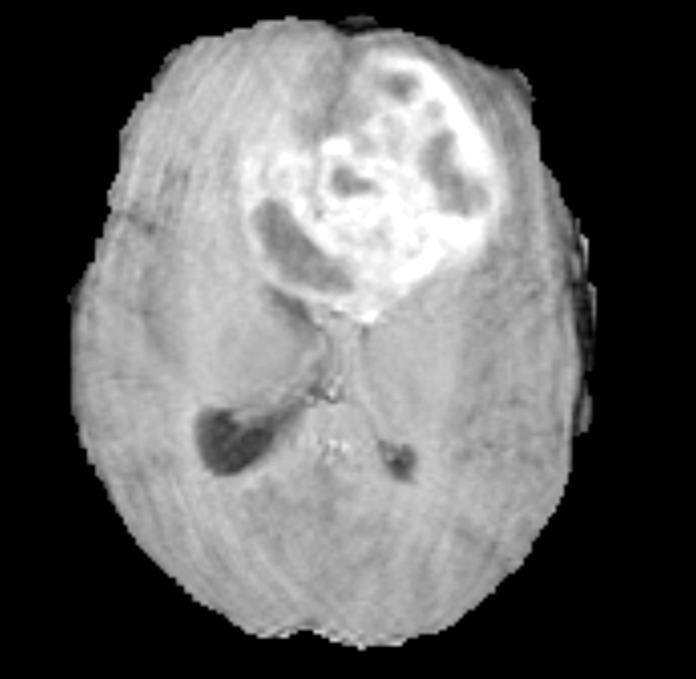}
        \caption{Butterfly}
        \label{fig:butterfly}
    \end{subfigure}
    \begin{subfigure}[t]{0.23\textwidth}
        \centering
        \includegraphics[width=4cm]{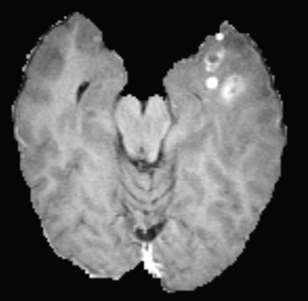}
        \caption{Multi-focal}       
        \label{fig:multifocal}
    \end{subfigure}
    \begin{subfigure}[t]{0.23\textwidth}
        \centering
        \includegraphics[width=4cm]{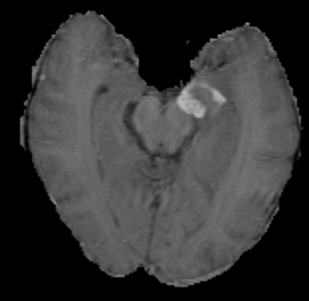}
        \caption{Non-convex}       
        \label{fig:noncovex}
    \end{subfigure}
    
    \caption{Examples of tumor shapes in the \textit{low-to-high} group. }  
    \label{fig:complex_tumor}
\end{figure}

We also examined the T1Gd images of two cases to explore the potential clinical implications. We observed that many of the \textit{low-to-high} cases correspond to well-recognized complex tumor shapes, such as butterfly glioblastoma (bGBM) and multi-focal glioma. Example scans of these shapes in the \textit{low-to-high} group are presented in Figure~\ref{fig:complex_tumor}. These shapes are known to be associated with aggressive progression and poor survival outcomes, as observed in bGBM \citep{dziurzynski2012butterfly,siddiqui2018butterfly,opoku2018surgical} and multifocal glioma \citep{patil2012prognosis,thomas2013incidence,di2019multiple}. 


\begin{figure}[!ht]
    \centering
    \begin{subfigure}[t]{0.23\textwidth}
        \centering
        \includegraphics[width=4cm]{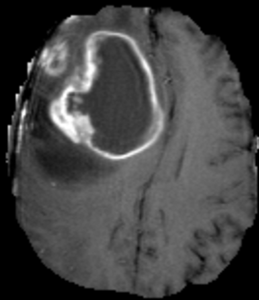}
    \end{subfigure}
    \begin{subfigure}[t]{0.23\textwidth}
        \centering
        \includegraphics[width=4cm]{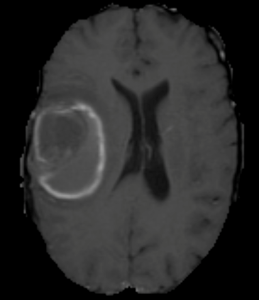}
    \end{subfigure}
    \caption{Examples of cystic regions.} 
    \label{fig:cystic_tumor}
\end{figure}



\begin{figure}[!ht]
\centering
    \includegraphics[width=17cm]{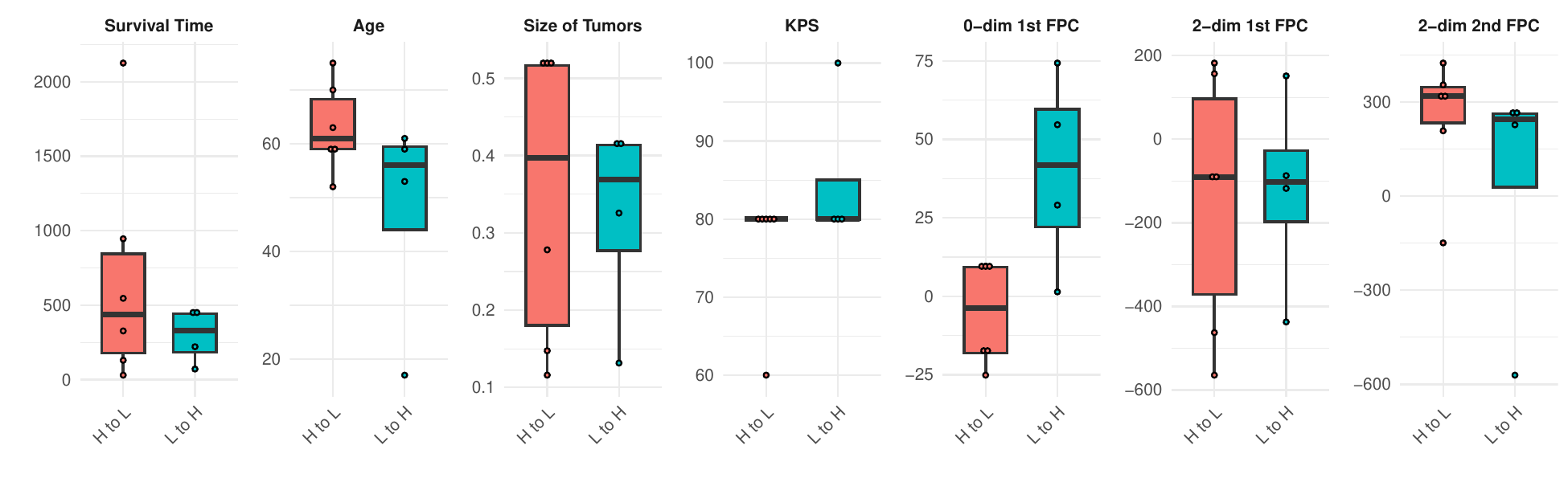}
    \caption{ Box plots of survival times, clinical variables, and FPC scores between the \textit{high-to-low} and \textit{low-to-high} groups, for the patients with cystic tumors.} 
    \label{fig:flip_cystic}
\end{figure}

We also observed ten cases with prominent cystic regions across both groups: six \textit{high-to-low} and four \textit{low-to-high} cases. The examples are given in Figure~\ref{fig:cystic_tumor}. Although these ten cases contain cystic regions that appear similar in certain slices, their overall shape characteristics differ between the groups. Figure~\ref{fig:flip_cystic} presents box plots summarizing the shape features of the cystic tumors, revealing group-wise differences in 0-dimensional and 2-dimensional topological features. This suggests that, despite local similarities in cystic regions, the proposed approach successfully captures global differences through comprehensive quantification of tumor shapes.

\section{Conclusion}
\label{sec:conc}
In this article, we proposed a predictive survival model for glioma patients, referred to as the PH-FCox model, which incorporates topological shape features extracted from segmented brain tumor images. Unlike the previous analysis by \cite{Crawford2020,Moon2023}, which utilized 2D slices of MRI scans with only two classes, our approach fully used 3D brain scans with three-class annotations, allowing for the extraction of richer shape information.

To obtain topological shape features, we apply the persistent homology method to the SEDT-3 values of segmented tumor images. The resulting persistence diagrams are smoothed with smoothing parameters determined by LOOCV, and subsequently incorporated into the PH-FCox model using the FPCA technique. The functional coefficients are estimated by maximizing the $\ell_1$-penalized partial likelihood, with the degree of smoothness selected through 10-fold cross-validation. 
By including interaction terms between shape features and tumor location, the model allows their effects to differ between frontal and non-frontal tumors.
Through the simulation study, we demonstrated that the proposed model successfully identifies shape differences and their lobe-specific effects on survival outcomes even under right-censoring.

The fitted PH-FCox model exhibited substantially improved predictive accuracy in classifying subjects into low- and high-risk groups based on the estimated predictive risks. We further examined GBM cases that were classified differently by the clinical-Cox and PH-FCox models to provide clinical insight. We found that the model effectively identified complex tumor morphologies and multifocal gliomas, both of which are known to be associated with poor survival outcomes. Even in cases exhibiting prominent cystic regions, the tumor shape characteristics remained predictive of survival outcomes.

\section*{Acknowledgments}
This research is supported by the National Cancer Institute (NCI) of the National Institutes of Health through grant R15 CA274241. 

\section*{Conflicts of Interest}
The authors declare no conflicts of interest.

\section*{Data Availability Statement}
The data that support the findings of this study are available from The Cancer Imaging Archive (TCIA). The code is available on \url{https://github.com/STATJANG/PH-FCox_model}.

\bibliographystyle{plain}
\bibliography{references}

\end{document}